\documentclass[prd,aps,superscriptaddress,nofootinbib,twocolumn,10pt]{revtex4-2}

\usepackage{graphicx} 
\usepackage[caption=false]{subfig}
\usepackage{amsmath} 
\usepackage{amssymb} 
\usepackage{physics}
\usepackage{hyperref}
\usepackage{xcolor}
\usepackage[capitalize]{cleveref}

\graphicspath{{figures/}}

\begin{document}

\title{Diamonds in the sky: the shape of gravitational waves from cosmic string cusps}
\author{Pierre Auclair}
\email{pierre.auclair@iap.fr}
\affiliation{Sorbonne Universit\'e, CNRS UMR 7095, Institut d'Astrophysique de Paris, 98 bis bd Arago, 75014 Paris, France}

\author{Chams Gharib}
\email{chams.gharib@phys.ens.psl.eu}
\affiliation{Sorbonne Universit\'e, CNRS UMR 7095, Institut d'Astrophysique de Paris, 98 bis bd Arago, 75014 Paris, France}
\affiliation{Laboratoire de Physique de l'ENS, Universit\'e Paris Cit\'e, Ecole normale sup\'erieure, Universit\'e PSL, Sorbonne Universit\'e, CNRS, 75005 Paris}

\author{Dani\`ele A.~Steer}
\email{daniele.steer@phys.ens.psl.eu}
\affiliation{Laboratoire de Physique de l'ENS, Universit\'e Paris Cit\'e, Ecole normale sup\'erieure, Universit\'e PSL, Sorbonne Universit\'e, CNRS, 75005 Paris}

\begin{abstract}
    We carry out an in-depth analysis of the gravitational wave (GW) emission from cosmic string cusps, going beyond the seminal work of \emph{Damour and Vilenkin}.
    We identify two features that, to the best of our knowledge, have not been described in the literature: first, the GW emission from a cusp, at any given frequency, is not concentrated inside a cone but instead draws a diamond-like shape in the sky.
    Second, the emission is not maximal in the direction of the cusp velocity, but in two slightly offset directions.
    We explain these effects in detail, and illustrate them with simplified examples.
    Then, we study two limiting cases for which the overall radiated power is enhanced: when the diamond is stretched in a particular direction, and when higher order derivatives vanish at the cusp.
    We statistically estimate the importance of these effects in a population of odd-harmonic loops.
\end{abstract}

\date{\today}

\maketitle

\section{Introduction}

Cosmic strings~\cite{Kibble:1976sj} are one-dimensional topological defects that may have formed in the early Universe during spontaneous symmetry breaking phase transitions.
They have an energy per unit length $\mu$ and relativistic velocities making them natural sources of gravitational waves (GWs).
In the one-dimensional thin-string limit, cosmic strings are described by the Nambu-Goto action~\cite{Nambu:1974zg}, and it is generically expected that cusps may appear on the cosmic string loops.
At a cusp, the string instantly reaches the speed of light, and several works in the eighties and nineties showed that cusps emit short duration, beamed, bursts of GWs~\cite{Vachaspati:1984gt, Vachaspati:1987sq, Allen:1991bk}.
Damour and Vilenkin \cite{Damour:2000wa, Damour_2001, Damour:2004kw} in particular were the first to determine the full cusp waveforms, see also Ref.~\cite{Siemens:2003ra}.

Since then, other works have extended these analytical results. 
For instance the gravitational waveform from kink-kink collisions was computed in Ref.~\cite{Binetruy:2010cc} (including the possible existence of Y-junctions on strings), and generalized in \cite{Ringeval:2017eww} for all types of burst-like events.
The emission of GW from cusps was studied numerically in Ref.~\cite{Blanco-Pillado:2017oxo} for ``smoothed'' cosmic string loops mimicking backreaction effects. Some of the tools introduced in that paper will be used here.
Precision modeling of the GW emission of cosmic string loops was also carried out in Ref.~\cite{Suresh:2023hkz}.

These different results have gained a new importance with the recent advent of GW astronomy. 
Indeed GW bursts from cosmic strings are actively searched for with the LIGO-Virgo-Kagra (LVK) ground-based GW detectors  ~\cite{LIGOScientific:2017ikf,LIGOScientific:2021nrg}, and they will also be searched for with future experiments such as the space-bound experiment LISA~\cite{Auclair:2023brk, Auclair:2023mhe, LISA:2024hlh} and Einstein Telescope (ET)~\cite{ET:2025xjr}.
The incoherent sum of these GW bursts (over all the lifetime of the cosmic string network) also generates as a Stochastic Gravitational Wave Background (SGWB). 
This is actively searched for in LVK~\cite{LIGOScientific:2017ikf,LIGOScientific:2021nrg,KAGRA:2025swt} and cosmic strings could provide an interpretation of the stochastic background detected by the different PTA collaborations around the globe~\cite{EPTA:2023xxk,NANOGrav:2023hvm,Zic:2023gta,Xu:2023wog}.
It is also one of the scientific objectives of LISA~\cite{Auclair:2019wcv,LISACosmologyWorkingGroup:2022jok,LISA:2024hlh,Blanco-Pillado:2024aca} and ET~\cite{ET:2025xjr}.

\begin{figure}[t]
    \includegraphics[width=.9\linewidth]{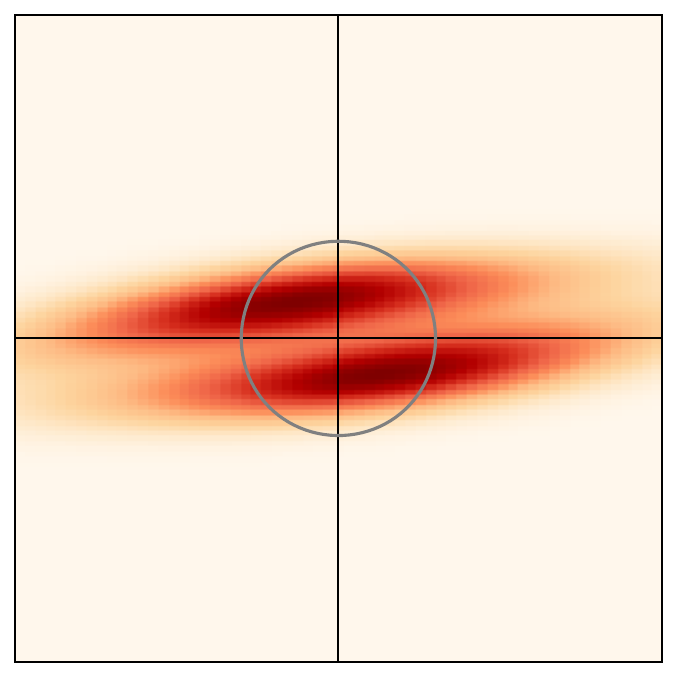}
    \caption{
        Projection onto the plane of $\abs{h}^\mathrm{TT}(\omega,\hat{n}) \equiv \sqrt{{h_{ij}^\mathrm{TT}}^*h_{ij}^\mathrm{TT}}$ for a given GW frequency $\omega$ and different observation directions $\hat{n}$. 
        In this figure the cusp feature sourcing the GWs points towards the center. 
        The colorscale goes linearly from $0$ (light) to the maxima (dark red) which occur when the observer is slightly misaligned with the cusp. 
        The circle shows the standard approximate shape often used in the literature. For more details on the exact configuration see \cref{fig:heatmap_hTT_beta36}.
    }
    \label{fig:detail_heatmap_hTT_beta36}
\end{figure}

Cusps are some of the most important and most studied features of cosmic string loops.
They are also known to emit particles of the field constituting the string~\cite{Blanco-Pillado:1998tyu, Olum:1998ag} which can modify the entire loop number density and thus the shape of the SGWB signal~\cite{Auclair:2019jip,Auclair:2021jud}, and have been proposed to source a nonlinear GW memory effect~\cite{Jenkins:2021kcj}.
They have been studied statistically in great detail in a population of high odd harmonic loops in Refs.~\cite{Pazouli:2020qmj, Pazouli:2021orp}, and for strings with Y-junctions~\cite{Elghozi:2014kya}.
Finally some extensions of the classical cusp scenario were proposed recently in Ref.~\cite{Drew:2025txu}.

In this paper, we revisit the seminal computation of the GW emission from a cusp of Refs.~\cite{Damour:2000wa,Damour_2001} and identify two features that, to the best of our knowledge, were not described before. 
Our focus is on the \emph{shape} of the GW beam emitted and its dependence on the GW frequency $\omega$.
First, we show that at any given $\omega$, the GW emission from a cusp is not concentrated inside a cone but instead draws a diamond-like shape in the sky.
Second, the direction of maximum emission at a given $\omega$ is not in the exact direction of the cusp velocity but is actually off-centered and concentrated in two opposing lumps around the cusp.
Both these features can be seen directly in \cref{fig:detail_heatmap_hTT_beta36}.

The purpose of this paper is to give a comprehensive explanation of these effects, and assess their possible impact on observational predictions.
In \cref{s:angular}, we revisit the standard computation of Refs.~\cite{Damour:2000wa,Damour_2001} and explain the origin of these two new features.
We show that for some simplified loop solutions the calculations can be done analytically, and that the GW emission can be substantially enhanced in certain special cases.
These include the `flat cusps' of Ref.~\cite{Drew:2025txu} for which we compute the beam aperture and overall GW power emission.
Finally in \cref{s:stats}, we carry out a statistical analysis to understand the prevalence of these enhanced configurations in a population of odd-harmonic loops, similarly to the analysis carried out in Refs.~\cite{Pazouli:2020qmj,Pazouli:2021orp}.
Concluding remarks are made in \cref{s:conclusion}.

\section{Angular structure of cusp radiation} \label{s:angular}

We first recall (\cref{sec:notation}) the basic properties of the cosmic string loops and the original computation of Ref.~\cite{Damour:2000wa} for the GR emitted by cusps. \cref{s:beamShape} explains how the non-conical structure of the beam emerges, and \cref{s:lumps} why the gravitational radiation is not maximal in the direction of the cusp.
We compute the effect on the overall radiated power in \cref{s:RadiatedPower}.
Finally, in \cref{s:FlatCusps}, we consider the generalized cusps recently discussed in Ref.~\cite{Drew:2025txu}.

\subsection{Notation and standard results}
\label{sec:notation}

\subsubsection{Setup}
\label{sec:setup}

In the thin-string limit, and working in the standard conformal-temporal gauge (see e.g.\ Ref.~\cite{Hindmarsh:1994re}), the string worldsheet $X^\mu(t,\sigma)$ is parametrized by time $t=X^0$ and a space-like coordinate $\sigma$ which measures equal energy intervals along the string.  It follows from the Nambu-Goto action in flat space that $X^\mu(t,\sigma)$ satisfies a wave equation~\cite{Vilenkin:2000jqa} whose solution is
\begin{equation}
    X^\mu(t, \sigma) = \frac{1}{2} \qty[X_+^\mu(\sigma_+) + X_-^\mu(\sigma_-)]
    \label{eq:bb}
\end{equation}
where $\sigma_\pm = t \pm \sigma$ and $X^0_{\pm} =\sigma_{\pm}$.
For closed loops in their center of mass frame, the focus of this paper, $-\ell/2 \leq \sigma \leq \ell/2$ where $\ell$ is the of invariant length of the loop, and the `loop halves' (also known as the left and right movers) $\vb{X}_\pm$ are periodic functions of period $\ell$. It then follows from \cref{eq:bb} that the loop itself has period $\ell/2$.  
To simplify notation, we introduce the usual dimensionless $2\pi$-periodic loop halves $\vb{a}(u)$ and $\vb{b}(v)$ through
\begin{equation*} 
    \vb{X}_+(\sigma_+) = \frac{\ell}{2\pi} \vb{a}\qty(u) ~~ \text{and} ~~ \vb{X}_-(\sigma_-) = \frac{\ell}{2\pi} \vb{b}\qty(v),
\end{equation*}
where 
\begin{equation}
u = 2\pi \frac{\sigma_+}{\ell}\, , \; \qquad v = 2\pi \frac{\sigma_-}{\ell} ,
\end{equation}
so that $-\pi\leq (u,v) \leq \pi$. The gauge conditions impose that the derivatives of the loop halves satisfy 
\begin{equation} \label{eq:Virasoro}
  \abs{\vb{a}'}^2=1=\abs{\vb{b}'}^2
\end{equation}
namely they are two closed curves on the unit Kibble-Turok sphere~\cite{KIBBLE1982141}. 
We also recall that cusps are points where $\vb{V} \equiv \pdv*{\vb{X}}{t}$ has modulus 1, namely 
\begin{equation}
    \vb{a}'= \vb{b}'. 
\label{eq:cusp}
\end{equation}
In general $\vb{X}''\neq 0$ at a cusp, though see Ref.~\cite{Drew:2025txu} and \cref{s:FlatCusps} below.

To first order in perturbation theory around Minkowski space (see e.g.\ Ref.~\cite{Speziale:2025zjp}) and working in the \emph{de Donder} gauge in the radiation zone, the Fourier transform of the trace-reversed metric perturbation 
\begin{equation}
    \bar{h}_{ij}(\vb{x}, \omega) = \int_{-\infty}^{+\infty} \dd{t} e^{i\omega t} \bar{h}_{ij}(\vb{x}, t),
\end{equation}
is related to the frequency domain stress energy tensor of the loop through
\begin{equation}
    \bar{h}_{ij}(\vb{x}, \omega) 
    = 4 G \frac{\ell}{2} 
    \frac{e^{i\omega \abs{\vb{x}}}}{\abs{\vb{x}}}T_{ij}(\vb{k}, \omega)
\end{equation}
where $\vb{k}$ satisfies the dispersion relation
\begin{equation}
\vb{k} = \omega \frac{\vb{x}}{\abs{\vb{x}}} \equiv \omega \hat{n}
\end{equation}
with $\hat{n}$ the unit vector pointing from the source to the observer, and
\begin{equation}
    T_{ij}(\vb{k}, \omega) = \frac{2}{\ell} \int_0^{\ell/2} \dd{t} \int \dd[3]{\vb{x}} e^{i(\omega t - \vb{k}\cdot\vb{x})}~T_{ij} (\vb{x}, t).
\end{equation} 
Furthermore, from the Nambu-Goto action (see e.g.\ Ref.~\cite{Vilenkin:2000jqa}) the stress-energy tensor takes the factorized form
\begin{equation} \label{eq:factorTmunu}
\begin{split}
    T^{ij}(k) =
    &\frac{\mu}{\ell} I^{(i}_+I^{j)}_- \\ 
    &\text{ with }
    \vb{I}_+(k) = \frac{\ell}{2\pi} \int_{-\pi}^{\pi} \dd{u} \vb{a}'(u) e^{ip\phi(u)}
\end{split}
\end{equation}
where $k^\mu = \omega(1,\hat{n})$, $\mu$ is the energy per unit length of the loop, and the dimensionless integer\footnote{This follows from the $\ell/2$-periodicity of the loop trajectory.} harmonic order $p$ is defined by
\begin{equation}
    p \equiv \frac{\ell \omega}{4 \pi}.
\end{equation}
Finally the phase is given by
\begin{equation} \label{eq:phi_of_u}
    \phi(u) \equiv u - \vb{a}(u)\cdot\hat{n}.
\end{equation}
The expression for $\vb{I}_-(k)$ is analogous with $\vb{a}'(u)$ replaced by $\vb{b}'(v)$.

Hence, determining the gravitational wave emission by a cosmic string loop amounts to calculating $\vb{I}_\pm$, from which
\begin{equation} \label{eq:hII}
    \bar{h}_{ij}(\vb{x}, \omega) = 2 G \mu \frac{e^{i \omega \abs{\vb{x}}}}{\abs{\vb{x}}} I^+_{(i}I^-_{j)}. 
\end{equation}
Then the two physical degrees of freedom
are obtained from transverse and traceless (TT) metric perturbation
\begin{equation}
\label{eq:TT}
    h_{ij}^\mathrm{TT} = \Lambda_{ij}^{~~k\ell}(\hat{n}) \bar{h}_{k\ell}
\end{equation}
where $\Lambda$ denotes the usual TT projector constructed from $\hat{n}$ \cite{Maggiore:2007ulw}.

\subsubsection{Cusps and beam width}

\begin{figure}[t]
    \centering
    \includegraphics[width=\linewidth]{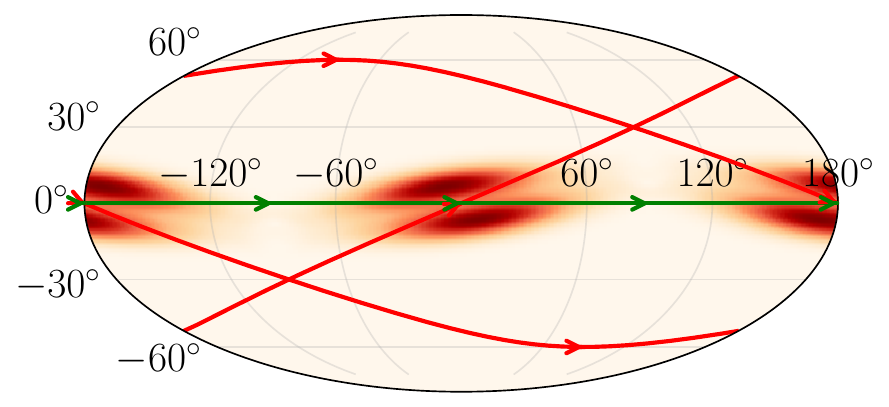}
    \caption{Mollenweide projection of the numerical calculation of $\abs{h}^\mathrm{TT}(\omega,\hat{n}) \equiv \sqrt{{h_{ij}^\mathrm{TT}}^*h_{ij}^\mathrm{TT}}$ as a function of the  observer's direction $\hat{n}$ . 
    The color scale goes linearly from $0$ (light) to maximum (dark red).
    We have superimposed the two curves $\mathcal{C}_{\vb{a'}}$ (red) and $\mathcal{C}_{\vb{b'}}$ (green), which correspond to a Kibble-Turok loop~\cite{Kibble:1982cb}.
    }
    \label{fig:heatmap_hTT_beta36}
\end{figure}

The integrals $\vb{I}_\pm$ appearing \cref{eq:factorTmunu} can be evaluated numerically, see \cref{fig:detail_heatmap_hTT_beta36}.
However, as discussed in Ref.~\cite{Damour_2001}, in the large frequency regime, they can also be evaluated using a saddle-point approximation.   We now recall these results as well as the beam-shape as calculated in Ref.~\cite{Damour_2001}.

From \cref{eq:phi_of_u}
the derivative of the phase is given by
\begin{equation}
    \phi'(u) = 1 - \vb{a}'(u) \cdot \hat{n} \geq 0 \, .
\end{equation}
Concerning \cref{eq:factorTmunu} there are thus two cases to distinguish in the large frequency limit $p \gg 1$. 
First, if $\hat{n}$ does not lie on the curve $\mathcal{C}_{\vb{a'}}$ drawn by $\vb{a}'$ on the Kibble-Turok sphere, then there is no real saddle point and the integral is exponentially suppressed at large frequencies. 
In the second case, if there exists $u^\star \in [-\pi, \pi]$ such that 
\begin{equation}
    \vb{a}'(u^\star) = \hat{n}
\end{equation}
then the integral $\vb{I}^+$ is dominated by this saddle point,
and it can be estimated as the sum of a longitudinal and a transverse parts with respect to $\hat{n}$
\begin{equation}
    \vb{I}_+(\hat{n}, p) \underset{p\rightarrow\infty}{\sim} \vb{I}_+^\parallel(\hat{n}, p) +  \vb{I}_+^\perp(\hat{n}, p) \, .
\end{equation}
These can be calculated in the standard way by expanding Taylor expanding the phase $\phi(u)$ as well as $\vb{a'(u)}$ around the saddle point (see e.g.\ Ref.~\cite{Damour:2000wa}), to yield
\begin{eqnarray}
    \frac{2\pi}{\ell}\vb{I}_+^\parallel 
    &=& \vb{a}'(u^\star) e^{i p \phi(u^\star)} 
    \qty[\frac{6}{p \vb{a}''^2(u^\star)}]^{1/3} 
    \frac{\Gamma(1/3)}{\sqrt{3}} \, ,
\nonumber
\\
\label{eq:IplusPerp}
    \frac{2\pi}{\ell}\vb{I}_+^\perp 
    &=& \vb{a}''(u^\star) e^{i p \phi(u^\star)} 
    \qty[\frac{6}{p \vb{a}''^2(u^\star)}]^{2/3} 
    \frac{i \Gamma(2/3)}{\sqrt{3}}.
\end{eqnarray}
The longitudinal term $\vb{I}_+^\parallel \propto \vb{a}'(u^*) = \hat{n}$ vanishes on applying the TT projector in \cref{eq:TT}.  
The transverse nature of $\vb{I}_+^\perp$ is guaranteed by $\vb{a}'\cdot \vb{a}'' = 0 $ which follows from $\vb{a}'^2 = 1$.
Thus $\vb{I}_+^\perp$
is the physically relevant quantity~\cite{Damour_2001}. 

The above derivation can be carried out identically for $\vb{I}_-(\hat{n}, p)$.
Thus from \cref{eq:hII}, the trace-reversed metric perturbation is exponentially suppressed at large frequencies in directions $\hat{n}$ that do not lie on both the curves $\mathcal{C}_{\vb{a'}}$ and $\mathcal{C}_{\vb{b'}}$ on the Kibble-Turok sphere. 
If, however, there exists $(u_c, v_c) \in [-\pi, \pi]^2$ such that 
\begin{equation}
    \hat{n} = \vb{a}'(u_c) = \vb{b}'(v_c)
\end{equation}
then the gravitational radiation is not exponentially suppressed. 
This corresponds to a 
cusp, see \cref{eq:cusp}, and the cusp radiates gravitational waves in the direction of its velocity $\vb{V}$ given by
\begin{equation}
    \vb{V} \equiv \eval{\pdv{\vb{X}}{t}}_{t_c, \sigma_c} = \vb{a}'(u_c) = \vb{b}'(v_c) = \hat{n}.
\end{equation}
From \cref{eq:hII,eq:IplusPerp}, the magnitude of the transverse and traceless metric perturbation scales as $p^{-4/3}$.

An obvious question to ask is how the GW emission decays when $\hat{n}$ is not exactly aligned with $\vb{V}$.
This was first addressed in Ref.~\cite{Damour_2001} who argued analytically --- using the saddle point approximation ---  that the beam's shape can be approximated by a cone with opening angle
\begin{equation} \label{eq:usualBeamWidth}
    \Delta \theta \propto p^{-1/3} .
\end{equation}

\begin{figure}[t]
    \includegraphics{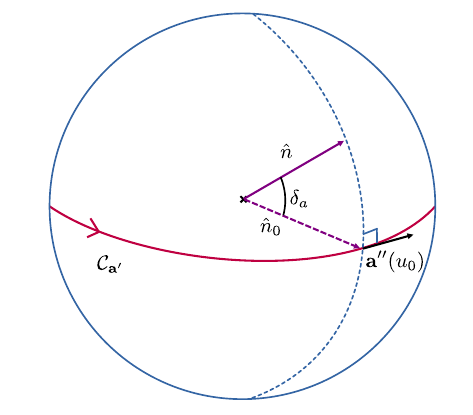}
    \caption{
    Misalignment between the observer's direction $\hat{n}$ and $\mathcal{C}_{\vb{a'}}$. Here we have shown $\mathcal{C}_{\vb{a'}}$ as a great circle, but this is purely for illustrative purposes. 
    }
    \label{fig:away_from_a_prime}
\end{figure}

\cref{fig:heatmap_hTT_beta36} shows our numerical calculation of
the magnitude of the transverse and traceless metric perturbation
\begin{equation}
    \abs{h}^\mathrm{TT}(\omega,\hat{n}) \equiv \sqrt{{h_{ij}^\mathrm{TT}}^*h_{ij}^\mathrm{TT}}
    \label{eq:habsdef}
\end{equation}
as a function of the direction of the observer, for a loop with $\mathcal{C}_{\vb{a'}}$ (red) and $\mathcal{C}_{\vb{b'}}$ (green) both describing a Kibble-Turok loop~\cite{KIBBLE1982141}, which intersect twice leading to 2 cusps.
(\cref{fig:detail_heatmap_hTT_beta36} shows detail of the central cusp, where the conical opening angle of \cref{eq:usualBeamWidth}, with proportionality factor 2, is also superimposed.) 
As the figure shows, most of the gravitational radiation is indeed emitted in the direction of the cusps.
However, we note two features that depart slightly from a conical structure. Indeed,
the shape of the beam does not display a circular symmetry; 
rather it is diamond-shaped, with axes determined by the tangent directions to $\mathcal{C}_{\vb{a}'}$ and $\mathcal{C}_{\vb{b}'}$ at the cusp (see \cref{s:angular}). 
Second, we observe that the maximum amplitude of the emitted gravitational radiation is not exactly in the direction of the cusp, but rather in two nearby directions, symmetrically distributed around the cusp. 

In the following we build an analytical understanding of these two features.

\subsection{Envelope of the beam}
 \label{s:beamShape}

To find the shape of the beam $\Delta\theta(\varphi)$, one must determine the exponential decay of $\vb{I}_+(\hat{n}, p)$ and $\vb{I}_-(\hat{n}, p)$ 
when $\hat{n}$, the direction of the observer, is slightly misaligned with the cusp direction $\vb{V}$. 

As a first step, consider the behavior of $\vb{I}_+(\hat{n}, p)$ when $\hat{n}$ lies slightly off the curve $\mathcal{C}_{\vb{a}'}$ on the Kibble-Turok sphere  (see \cref{fig:away_from_a_prime} for details). 
Let $\hat{n}_0$ denote a reference direction on $\mathcal{C}_{\vb{a'}}$ (not necessarily a cusp); 
\begin{equation}
   \hat{n}_0 = \vb{a}'(u_0)  \quad \text{with} \quad u_0 \in [-\pi, \pi].
\end{equation}
We obtain $\hat{n}$ by rotating $\hat{n}_0$ with respect to $\vb{a}''(u_0)$\footnote{Recall that, since $\abs{\vb{a}'} = \abs{\vb{b'}} = 1$, the first and second derivatives of the loop halves are always perpendicular (i.e.\ $\vb{a}'\cdot \vb{a}'' = 0$) and can be used to define a local basis.} by an angle $0 \leq \delta_a \ll 1$. 
To estimate $\vb{I}_+$ in the large frequency regime, we look for the \emph{complex roots} $u^\star$ of $\phi'(u)=1-\vb{a}'(u)\cdot \hat{n}$ to carry the saddle point approximation. More precisely, see Appendix~\ref{s:saddlePoint}, the root for which $\phi''(u^*)$ has a \emph{positive} imaginary component will drive the exponential decay of the integral. 
Taylor expanding around $u_0$ gives
\begin{equation}
\begin{split}
    0 = \phi'(u^\star) &= \phi'(u_0)\\ &+ (u^\star - u_0) \phi''(u_0)\\ &+ \frac{1}{2} (u^\star -u_0)^2 \phi^{(3)}(u_0) + \mathcal{O}(\delta_a^3)
\end{split}
\label{eq:split}
\end{equation}
where we have assumed that
\begin{equation*}
    u^\star = u_0 + \order{\delta_a},
\end{equation*}
and where, as $\hat{n}_0$ lies on $\mathcal{C}_{\vb{a'}}$,
\begin{eqnarray}
\label{eq:derivativesAtu0}
    \phi'(u_0) &=& 1 -\cos \delta_a = \frac{\delta_a^2}{2} + \order{\delta_a^4}, 
    \\
    \phi''(u_0) &=& 0,
\label{eq:derivativesAtu0Bis}
\\
    \phi^{(3)}(u_0) &=& \qty[\vb{a}''(u_0)]^2 + \order{\delta_a} \, .
\end{eqnarray} 
Thus from \cref{eq:split}
\begin{equation}
    u^\star = u_0 + i \frac{\delta_a}{\abs{\vb{a}''(u_0)}} + \order{\delta_a^{2}}.
\end{equation}
In the large frequency limit, the integral $\vb{I}_+$ can therefore be estimated by deforming the contour of integration in the complex plane and going through $u^\star$ to yield (see Appendix~\ref{s:saddlePoint} for further details)
\begin{equation} \label{eq:decayI}
    \abs{\vb{I}_+(\hat{n}, p)} \propto \abs{e^{i p \phi(u^\star)}} = 
    \exp\qty[-\frac{1}{2} \left(\frac{\delta_a}{\Delta_a(u_0,p)} \right)^3]    
\end{equation}
where
\begin{equation}
    \Delta_a(u_0, p) \equiv \qty[\frac{\abs{\vb{a}''(u_0)}}{p}]^{1/3}.
\end{equation}

The above derivation proceeds in a similar way for $\vb{I_-}$.
\begin{figure}
    \centering
    \includegraphics{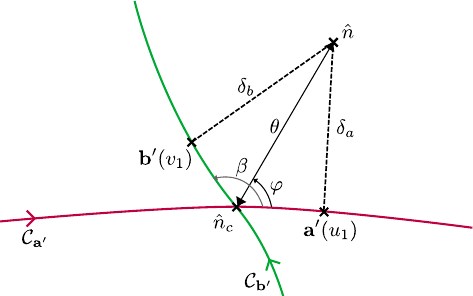}
    \caption{
        Representation of relevant geometrical quantities in the plane tangent to the Kibble-Turok sphere at the cusp. 
        The distances $\delta_{a, b}$ can be evaluated by approximating $\mathcal{C}_{\vb{a}', \vb{b}'}$ by their tangent at the cusp
    }
    \label{fig:delta_a_b}
\end{figure}
Then on combining the two one can determine the decay of the GW amplitude for  observer direction $\hat{n}$ slightly misaligned with a cusp at $(u_c, v_c)$ where
\begin{equation}
 \vb{V}\equiv   \hat{n}_c = \vb{a}'(u_c) = \vb{b}'(v_c).
\end{equation}
Indeed, using \cref{eq:hII,eq:decayI}, the trace-reversed metric perturbation is controlled by
\begin{equation}
    \abs{\bar{h}_{ij}} \propto \exp\left\{-\frac{1}{2} \left[
\qty(\frac{\delta_a}{\Delta_a(u_1, p)})^3 + \qty(\frac{\delta_b}{\Delta_b(v_1, p)})^3
\right] \right\}
    \label{eq:awful}
\end{equation}
where $\delta_a$ (resp.~$\delta_b$) is the angular distance of $\hat{n}$ to $\mathcal{C}_{\vb{a'}}$ (resp.~$\mathcal{C}_{\vb{b'}}$) on the Kibble-Turok sphere, and $u_1$ (resp.~$v_1$) is the coordinate corresponding to the projection of $\hat{n}$ on $\mathcal{C}_{\vb{a'}}$ (resp.~$\mathcal{C}_{\vb{b'}}$), see \cref{fig:delta_a_b} 
\footnote{
This result is consistent with the expressions of Ref.~\cite{Blanco-Pillado:2017oxo} however they did not comment on the beam shape, nor on the misaligned maxima of the amplitude. }.
In this form, \cref{eq:awful} contains all the information we need to understand the diamond-like shape of the cusp emission. 
The emission is not isotropic around the cusp direction but is wrapped in two ribbons that follow $\mathcal{C}_{\vb{a'}}$ and $\mathcal{C}_{\vb{b'}}$.

\begin{figure}[t]
    \includegraphics[width=6cm]{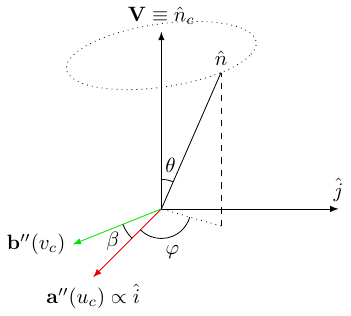}
    \caption{
    Representation of relevant geometrical quantities in three dimensions at the cusp. 
    }
    \label{fig:Dani}
\end{figure}

However, the more physically relevant parameters are, see \cref{fig:Dani}:
\begin{itemize}
    \item $\theta \ll 1$, the angle between $\hat{n}$ and $\hat{n}_c$
    \item $\varphi \in [0,2\pi]$, the rotation angle of $\hat{n}$ around $\hat{n}_c$
    \item $\beta$, the angle between $\vb{a}''_c$ and $\vb{b}''_c$
    \item the successive derivatives of $\vb{a}_c'$ and $\vb{b}_c'$.
\end{itemize}
Then, using \cref{fig:delta_a_b,fig:Dani} (see Appendix~\ref{app:demo} for further details), the relevant terms in $\delta_a$ are given by
\begin{equation}\label{eq:delta_a_full}
    \delta_a^2 \approx \theta^2 \sin^2 \varphi  
    + q_a^2 \theta^4 \cos^4 \varphi 
\end{equation}
where the geometrical factor $q_a$ is given by
\begin{equation}
   q_a = \frac{1}{2} \left[\frac{\vb{a}^{(3)}(u_c)\cdot \hat{j}}{\abs{\vb{a''}(u_c)}^2}\right].
\end{equation}
Here the third derivative $\vb{a}^{(3)}(u_c)$ describes the curvature of $\mathcal{C}_{\vb{a}'}$ around the cusp, and we have defined a triad around the cusp as (see \cref{fig:Dani})
\begin{equation}
    \hat{i} \equiv \frac{\vb{a}''(u_c)}{\abs{\vb{a}''(u_c)}}, \quad
    \hat{j} \equiv \vb{a}'(u_c) \cross \hat{i}.
    \label{eq:basis}
\end{equation}
A similar expression holds for $\delta_b$.

\begin{figure*}[t]
    \subfloat[$p = 300$]{
        \includegraphics[width=.48\textwidth]{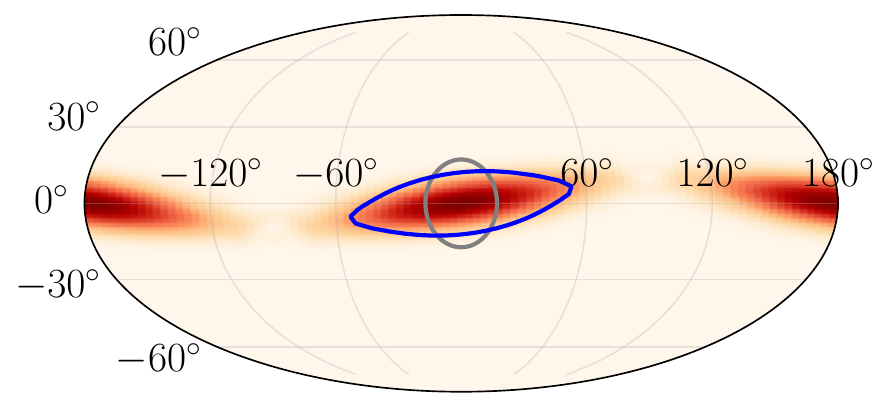}
    }
    \subfloat[$p = 3000$]{
        \includegraphics[width=.48\textwidth]{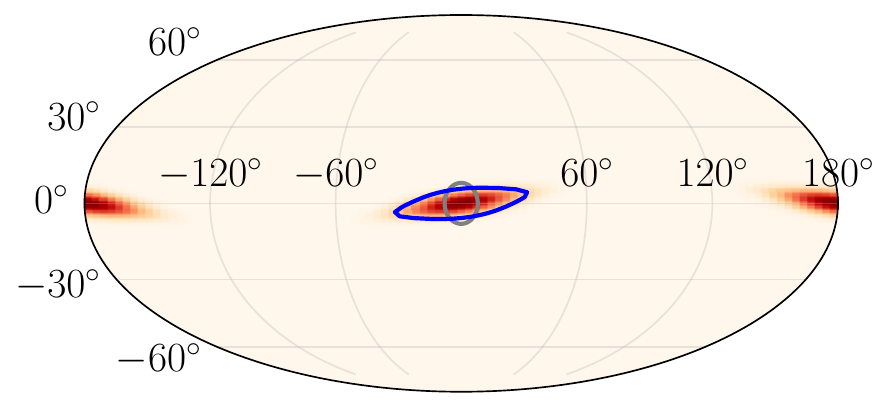}
    }
    \caption{
        Mollenweide projection of the numerical calculation of the trace-reversed metric $\sqrt{{\bar{h}_{ij}}^*\bar{h}_{ij}}$ of \cref{eq:hII} as a function of $\hat{n}$, for different values of $p$.
        The beam aperture formula of \cref{eq:beamWidth} is shown in blue.
        The standard formula \cref{eq:usualBeamWidth} of Ref.~\cite{Damour_2001}, with a proportionality constant of $2$ is shown in gray.  
    }
\label{fig:heatmap_hTT_beamwidth}
\end{figure*}

If the first term in $\delta_a, \delta_b$ is non-vanishing, i.e.\ provided $\sin\varphi \neq 0$ or $\sin(\varphi-\beta) \neq 0$, then
\cref{eq:awful} becomes
\begin{equation} \label{eq:expDecayh}
    \abs{\bar{h}_{ij}} \propto \exp{-4 \qty[\frac{\theta}{\Delta\theta(\varphi)}]^3}
\end{equation}
where the typical beaming angle $\Delta\theta$ depends on the precession angle $\varphi$ through
\begin{equation}\label{eq:beamWidth}
    \Delta \theta(\varphi) \approx \frac{2}{p^{1/3}} \qty[\frac{\abs{\sin\varphi}^3}{\abs{\vb{a}''(u_c)}} + \frac{\abs{\sin(\varphi-\beta)}^3}{\abs{\vb{b}''(v_c)}}]^{-1/3} .
\end{equation}
(Note that we have factored out a factor 4 in \cref{eq:beamWidth}, for convenience, in order to capture the typical decay width of the amplitude.) 
To check the validity of \cref{eq:beamWidth} we numerically integrated the integrals $\vb{I}_+$ and $\vb{I}_-$ using different (analytic) $\vb{a}$ and $\vb{b}$ leading to cusps, and finally compared their decay with \cref{eq:beamWidth}. 
As can be seen in \cref{fig:heatmap_hTT_beamwidth}, \cref{eq:beamWidth} correctly captures the angular decay of the gravitational radiation around the cusp.

At this point it is useful to note the following. 
First, due to the dependence of the beam width on $\varphi$, the beam shape is \emph{not} conical, but rather a `diamond-like', and extended in the directions of $\vb{a}''(u_c)$ and $\vb{b}''(v_c)$. 
This anisotropy of the beam width was not discussed in the original computation of Ref.~\cite{Damour_2001}, as the term causing it was in fact discarded, see Appendix~\ref{app:deviation}. 
Second, \cref{eq:beamWidth} shows the same $p^{-1/3} \sim (\omega \ell)^{-1/3}$ behavior of the beam width as in Ref.~\cite{Damour_2001} (c.f.~\cref{eq:usualBeamWidth}), and hence gravitational radiation from a cusp is concentrated in a narrow beam directed by the velocity of the cusp. 

If the first term in \cref{eq:delta_a_full} vanishes (and similarly for $\delta_b$), that is in the limit $\varphi, \beta \to 0$, then 
\begin{equation}
    \delta_a  \underset{\varphi = 0}{=}q_a \theta^2 \, .
\end{equation}
Therefore \cref{eq:awful} becomes
\begin{equation} \label{eq:expDecayh}
    \abs{\bar{h}_{ij}} \underset{\varphi = 0}{\propto} \exp{-4 \qty[\frac{\theta}{\Delta\theta_{{\beta=\varphi=0}}}]^6}
\end{equation}
where the finite beam aperture is now given by
\begin{equation} \label{eq:deltaThetaBetaZero}
    \Delta \theta_{\beta = \varphi= 0} = \sqrt{2} \qty[ \frac{\abs{\vb{a}''(u_c)}}{q_a^{3}} + \frac{\abs{\vb{b}''(u_c)}}{q_b^{3}}]^{1/6} p^{-1/6}.
\end{equation}
The beam aperture is therefore enhanced from the standard scaling in $p^{-1/3}$ to broader $p^{-1/6}$.
In Appendix~\ref{app:demo} we discuss this in more detail and show that corrections to the beam aperture given in \cref{eq:beamWidth} need to be taken into account when $\beta$ is smaller than a threshold value
\begin{equation} \label{eq:betaThreshold}
    \beta_\text{th} \propto \qty(\frac{1}{p})^{1/6}.
\end{equation}

The modification of the beam width might have several observational consequences. 
Indeed, the enhancement provided either by the diamond like shape, or by the change in the asymptotic behavior as $\beta \to 0$ could make GW bursts more likely to be observed by a distant observer, and therefore increase the expected rate of detections in burst searches~\cite{LIGOScientific:2017ikf,LIGOScientific:2021nrg,Auclair:2023brk,Auclair:2023mhe}. 
Moreover, the enhancement of the beam aperture could have an impact on the total radiated power, thus modifying the predictions for the SGWB~\cite{LIGOScientific:2021nrg,LISA:2024hlh,ET:2025xjr}. 
We delay the evaluation of the radiated power to \cref{s:RadiatedPower}, and focus first on explaining the two lumps of gravitational radiation observed in \cref{fig:detail_heatmap_hTT_beta36,fig:heatmap_hTT_beta36}.

\subsection{Directions of maximal emission} \label{s:lumps}
\begin{figure*}[t]
    \centering
    \subfloat[$\beta = \pi/2$, $p = 100$ \label{fig:eps=+1}]{
        \includegraphics[width=.48\textwidth]{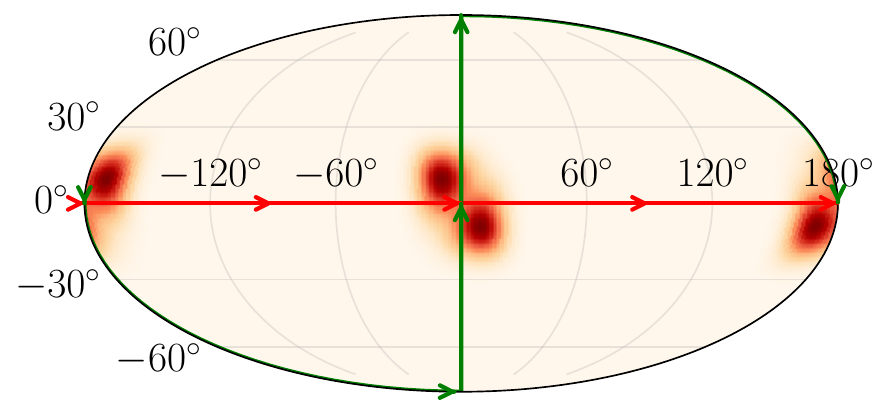}
    }
    \hfill
    \subfloat[$\beta = -\pi/2$, $p = 100$ \label{fig:eps=-1}]{
        \includegraphics[width=.48\textwidth]{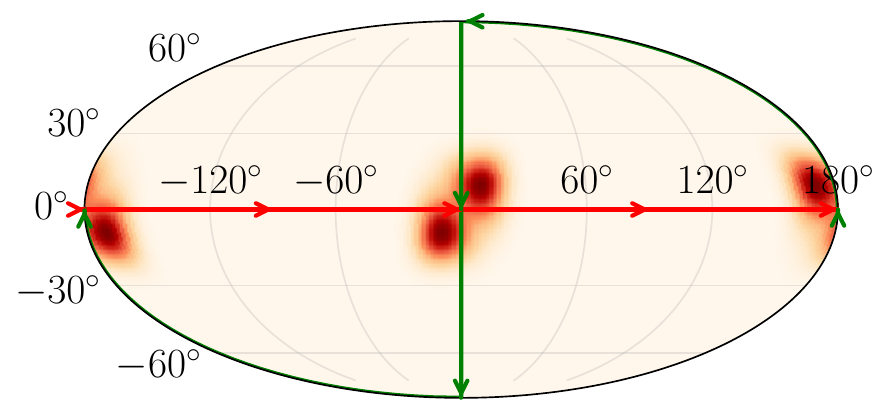}
    }
    \caption{Numerical calculation of $\abs{h}^\mathrm{TT}(\omega,\hat{n})$ with $\vb{a'}(u)$ (red) and $\vb{b}'(v)$ (green) chosen as orthogonal great circles. The arrows indicate the increasing directions of $u$ and $v$.}
    \label{fig:heatMaps_TT_effects}
\end{figure*}

As observed in e.g.\ \cref{fig:detail_heatmap_hTT_beta36,fig:heatmap_hTT_beta36}, the numerical calculation of $\abs{h}^\mathrm{TT}$ shows two peaks in directions symmetrically distributed around the cusp.  
Furthermore, as \cref{fig:heatMaps_TT_effects} shows, the positions of these peaks depends on the orientation of the curves $\mathcal{C}_{\vb{a}', \vb{b}'}$. 
As we now discuss, this effect comes from the transverse-traceless projection.  

\begin{figure}[t!]
    \centering  \includegraphics{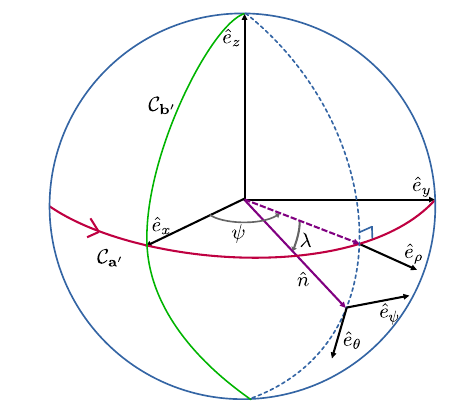}
    \caption{
        Geometrical parametrization used in \cref{s:lumps} for describing the direction of the observer $\hat{n}$ on the Kibble-Turok sphere, with $\mathcal{C}_{\vb{a}'}$ being a great circle at the equator.
    }   
    \label{fig:geom_exact_calc}
\end{figure}

To understand this, consider a simplified --- and analytically soluble --- case 
in which $\mathcal{C}_{\vb{a'}}$ and $\mathcal{C}_{\vb{b'}}$ are two orthogonal great circles on the Kibble-Turok sphere. (This calculation can be straightforwardly generalized to any $\beta \neq \pm \pi/2$.)
We place $\mathcal{C}_{\vb{a'}}$ on the equator
\begin{equation}
\label{eq:sundaynight}
    \vb{a}'(u) = \cos u \, \hat{e}_x + \sin u \, \hat{e}_y,
\end{equation}
and the observer's direction $\hat{n}$
is characterized by the angles $\lambda$ and $\psi$ in spherical coordinates, see \cref{fig:geom_exact_calc}.
Then the phase $\phi(u)$ in \cref{eq:phi_of_u} becomes
\begin{equation}
    e^{ip\phi(u)} = e^{ipu} e^{-ip\cos\lambda \sin(u-\psi)}\, 
\end{equation}
where the second term can be written as \cite{Gradshteyn:1702455}
\begin{equation}
    e^{-ip \cos\lambda\sin(u-\psi)} = \sum_{n=-\infty}^{n=+\infty} J_n(p\cos\lambda) e^{-in(u-\psi)}
\end{equation}
with $J_n$ is the $n$-th Bessel function of the first kind. Substituting \cref{eq:sundaynight} into $\vb{I}_+$ (see \cref{eq:factorTmunu}), and integrating over $u$ gives
\begin{equation} \label{eq:IplusGreatCircles}
    \vb{I}_+ = \frac{\ell}{2\pi} e^{ip\lambda}\qty[I_\rho(p,\lambda) \hat{e}_\rho + I_\psi(p,\lambda) \hat{e}_\psi]
\end{equation}
where the basis vectors are directed in the radial ($\rho$) and orthogonal ($\psi$) direction
\begin{equation}
    \begin{cases}
    \hat{e}_\rho &= \cos\psi \, \hat{e}_x + \sin\psi \, \hat{e}_y \\
    \hat{e}_\psi &= -\sin\psi \, \hat{e}_x + \cos\psi \, \hat{e}_y,
    \end{cases}
\end{equation}
and where \begin{equation}
    \begin{cases}
        I_\rho(p,\lambda) &= \pi \qty[ J_{p-1}(p\cos\lambda) + J_{p+1}(p\cos\lambda) ] \\
        I_\psi(p,\lambda) &= i \pi \qty[ J_{p-1}(p\cos\lambda) - J_{p+1}(p\cos\lambda) ] \, ,
    \end{cases}
    \label{eq:is}
\end{equation}
with $I_\psi(p,\lambda) $ is purely imaginary.
As expected, since the loop radiates GWs only at harmonics of the fundamental frequency $2/\ell$, $\vb{I}_+$ in \cref{eq:IplusGreatCircles} vanishes when $p$ is non-integer. 

We place $\mathcal{C}_{\vb{b'}}$ in the $(\hat{e}_x,\hat{e}_z)$-plane, see \cref{fig:geom_exact_calc}, so that the cusp is at $\psi_c = \lambda_c=0$. 
The expression for $\vb{I}^-$ is as in \cref{eq:IplusGreatCircles} though with the vectors $\hat{e}_\rho$ and $\hat{e}_\psi$, as well as the angles $\psi$ and $\lambda$ straightforwardly adapted to the great circle spanned by $\mathcal{C}_{\vb{b'}}$.

When $\hat{n}$ is in the direction of the cusp, the radial ($\rho$) component is longitudinal.
Therefore, the radial component disappears in the TT gauge, and the $\psi$ component is the physical one.
This is not the case anymore away from the great circle, as $\hat{e}_\rho$ and $\hat{n}$ become misaligned, as in \cref{fig:geom_exact_calc}.
The radial component receives a transverse part, thereby adding a contribution to the gravitational wave amplitude away from the cusp, making it actually peak in a slightly displaced direction. 

To see this in more detail, let
\begin{equation}
    \xi = \begin{cases}
        +1 & \text{if } \beta = +\pi/2  \\
        -1 & \text{if } \beta = -\pi/2
    \end{cases}
\end{equation}
and now focus on the angular behavior of $h_{ij}^\mathrm{TT}$, that is we ignore the prefactors of \cref{eq:hII}.
For a given $\hat{n}$, the two unit vectors $\hat{e}_\lambda$ and $\hat{e}_\psi$ can be used to build the $+$ and $\times$ transverse and traceless polarization tensors
\begin{equation}
    \begin{cases}
        e^+_{ij} &= {e_\lambda}_i {e_\lambda}_j - {e_\psi}_i {e_\psi}_j \\
        e^\times_{ij} &= {e_\lambda}_i {e_\psi}_j + {e_\lambda}_j {e_\psi}_i
    \end{cases}
\end{equation}
from which the $+$ and $\times$ amplitudes of the GW waves can be extracted in the usual way $2 h_{+, \times} = \bar{h}_{ij} e^{+, \times}_{ij}$
with
\begin{equation}
    \abs{h}^\mathrm{TT}
    = \sqrt{2} \sqrt{\abs{h_+}^2 + \abs{h_\times}^2}.
\end{equation}
Using $\bar{h}_{ij} \propto I^+_{(i} I^-_{j)}$, we find 
\footnote{Here and in the following, we do not report the prefactor $e^{i p\varphi}$ appearing in \cref{eq:IplusGreatCircles} as it does not affect the derivation and eventually disappears when calculating $\abs{h_{+, \times}}^2$.}
\begin{equation}
    \begin{cases}
        2 h_+ &= \qty[ (\vb{I}^+ \cdot \hat{e}_\lambda)(\vb{I}^- \cdot \hat{e}_\lambda) - (\vb{I}^+ \cdot \hat{e}_\psi)(\vb{I}^- \cdot \hat{e}_\psi) ] \\
        2 h_\times &= \qty[ (\vb{I}^+ \cdot \hat{e}_\lambda)(\vb{I}^- \cdot \hat{e}_\psi) + (\vb{I}^+ \cdot \hat{e}_\psi)(\vb{I}^- \cdot \hat{e}_\lambda) ] \, .
    \end{cases}
\end{equation}

As expected, $h_+$ vanishes at the cusp (that is at $\lambda =\psi = 0$)
since in this orthogonal configuration, $\vb{I^+} \cdot \hat{e}_\lambda = \vb{I^-} \cdot \hat{e}_\psi = 0$.
Indeed, it is known that the gravitational waves radiated by a cusp are linearly polarized~\cite{Damour_2001}.
In the following, we focus on the behavior of $\abs{h_\times}^2$ near the cusp, as it is enough to understand the location of the maxima of the GW amplitude. 

To show that the cusp does not correspond to the maximum of the GW amplitude, we expand the GW strain to second order in small angles ($\psi, \lambda$) around the cusp.
To lowest order,
\begin{equation} \label{eq:hPlusOrdre1}
    \begin{cases}
        2 h_+ &= I_\psi(p,0) I_\rho(p,0)( \psi + \xi \lambda) + \order{2} \\
        2 h_\times &= -\xi I^2_\psi(p,0) + \order{1} \, .
    \end{cases}   
\end{equation}
Recall that $I_\rho$ and $I_\psi$ are given in \cref{eq:is} and that $I_\psi$ is purely imaginary.
Hence, $h_+ / h_\times \in i \mathbb{R}$, showing that the polarization is no longer linear but elliptical.
Indeed, there does not exist a rotation of angle $\alpha$ that can make $h_+$ vanish, i.e.\ such that $\tan 2\alpha = -h_+ / h_\times$ since it is a purely imaginary number.

Expanding $\abs{h_\times}^2$ to second order gives
\footnote{
    A similar analysis can be performed for the subdominant $\abs{h_+}^2$ polarization by taking the squared modulus of \cref{eq:hPlusOrdre1}, which yields the eigenvalues $1 \pm \xi$, positive in the growing direction of $\abs{h_\times}^2$. 
}
\begin{multline}
    4 \abs{h_\times}^2 = 
    I^4_\psi(p,0)  \\
    + I^2_\psi(p,0) 
        \mqty(\lambda \\ \psi)^\text{T} 
        \mqty( I_\psi(p,0) I_\psi''(p,0) & -\xi I^2_\rho(0) \\ -\xi I^2_\rho(p,0) & I_\psi(p,0) I_\psi''(p,0)) 
        \mqty(\lambda \\ \psi) \\
        + \order{3}.
\end{multline}
The anisotropy comes from the 2nd term which has eigenvectors $\lambda = \pm \varphi$, with associated eigenvalues
\begin{equation}
    \alpha_\pm = \frac{1}{4} I^2_\psi(p,0) \qty[I_\varphi(p,0) I_\psi''(p,0) \mp \xi I^2_\rho(p,0)].
\end{equation}
This helps explain the two panels of \cref{fig:heatMaps_TT_effects}.
In \cref{fig:eps=+1}, $\beta = \pi/2$, so $\xi=1$.
Then $\alpha_+ > 0$ and $\alpha_- < 0$, so the axis carrying the maxima of the amplitude is given by ${\lambda} = \varphi$, i.e.\ the top-left and bottom-right corners of \cref{fig:eps=+1} as observed there. 
On the other hand for \cref{fig:eps=-1}, $\beta = -\pi/2$ so $ \xi = -1$, and the maxima are on the axis given by $\lambda = -\varphi$, i.e.\ the top-right and bottom-left corners.

\begin{figure}
    \centering
    \includegraphics{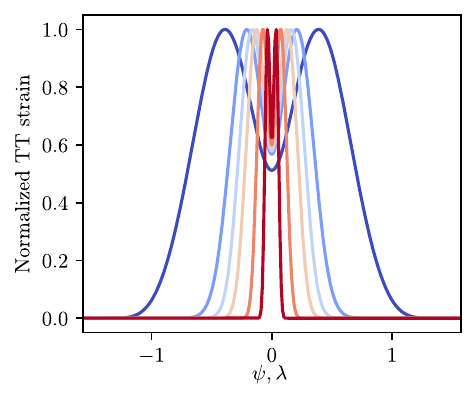}
    \caption{Profile of $\abs{h}^\mathrm{TT}$ (normalized) near the cusp along the axis defined by $\psi = \lambda$, for different values of $p$ ($10$, $50$, $100$, $200$, $1000$, $10~000$). Higher values of $p$ are red, while lower are blue. Here $\beta = \pi/2$.}
    \label{fig:lump_amplitude}
\end{figure}

Quantitatively, \cref{fig:lump_amplitude} shows that the ratio of the amplitude of the lumps over the amplitude at the cusp is of order unity. 
This explains why this effect could have been neglected in previous work (it does not appear for example plotting on a logarithmic scale).
Moreover, this shows that the amplitude of the GW is well approximated by its value at the cusp, even if it is not exactly the maximum.

As expected, we see in the definition of the $\alpha_\pm$ that the transverse and traceless projection generates an interference phenomenon between the radial ($\rho$) and orthogonal ($\psi$) components. 
There are two preferred axes that are equidistant to the great circles, one of which carries the maxima of the GW amplitude. 

In general, beyond this simple configuration of 2 great circles, the lumps in the diamond shape are delineated by $\vb{a}''(u_c)$ and $\vb{b}''(v_c)$. This is particularly clear in \cref{fig:heatmap_hTT_beta36}.

\subsection{Radiated power} \label{s:RadiatedPower}

In this section we estimate the power radiated in GWs, taking account the  beam shape discussed above.
The trace reversed metric perturbation can be approximated by
\begin{equation}
    \bar{h}_{ij}(\vb{x}, \omega) \approx 2 G\mu \frac{e^{i \omega \abs{\vb{x}}}}{\abs{\vb{x}}} \qty[I^+_{(i} I^-_{j)}]_{\hat{n} = \hat{n}_c} \Theta_\mathrm{H}\qty[\Delta \theta(\varphi) - \theta]
\end{equation}
where $\Theta_\mathrm{H}$ denotes the Heaviside step function, and where (as in  \cref{fig:Dani}), $\theta \ll 1$ and $\varphi \in [0, 2\pi]$ parametrize the misalignment between $\hat{n}$ and $\hat{n}_c$. 
The magnitude of the transverse and traceless metric perturbation scales as  (see \cref{eq:IplusPerp})
\begin{equation}\label{eq:httx}
    {h_{ij}^\mathrm{TT}(\vb{x}, \omega)} \propto 2\frac{G \mu}{\abs{\vb{x}}} \left( \frac{\ell}{2\pi}\right)^{2/3} \frac{\Theta_\mathrm{H}\qty[\Delta \theta (\varphi) - \theta]}{\qty(\omega^4 \abs{\vb{a}''(u_c)} \abs{\vb{b}''(v_c)})^{1/3}} .
\end{equation}
Note that considering the amplitude at the cusp is justified since, as mentioned at the end of \cref{s:lumps}, the ratio of the amplitude at the maxima over the amplitude at the cusp is $\order{1}$.

Recalling the usual formula for the radiated power
\begin{equation}
    P = \frac{1}{32 \pi G} \int \dd{\Omega} r^2 \expval{\dot{h}_{ij}^\mathrm{TT} \dot{h}_{ij}^\mathrm{TT}},
\end{equation}
the radiated power per unit frequency in the large frequency regime can be estimated by
\begin{equation} \label{eq:dpdwWithAngle}
    \dv{P}{\omega} \approx G \mu^2 \frac{\ell^{1/3}}{\omega^{2/3}} \int_0^{2\pi} \dd{\varphi} \Delta \theta(\varphi)^2.
\end{equation}
If one assumes --- as is usually done in the literature --- that $\Delta \theta$ is independent of $\varphi$ and given by \cref{eq:usualBeamWidth}, this leads to
\begin{equation} \label{eq:dpdwusual}
    \dv{P}{\omega} \approx G \mu^2 \frac{1}{\ell^{1/3} \omega^{4/3}} \, ,
\end{equation}
which is consistent with equations (A.9) and (A.10) of Ref.~\cite{Auclair:2019wcv}.
However, as we have seen above, the emission is not exactly in a cone and rather draws a diamond-like shape in the sky described by \cref{eq:beamWidth}.
As an order of magnitude estimation, we approximate this diamond by a rhombus, namely we simplify $\Delta \theta$ as 
\begin{equation}
    \Delta \theta \approx \qty(\frac{1}{\ell \omega})^{1/3} \min\qty[ \frac{2 \pi\abs{\vb{a}''(u_c)}}{\abs{\sin\varphi}},  \frac{2\pi \abs{\vb{b}''(v_c)}}{\abs{\sin(\varphi - \beta)}}]
\end{equation}
which, according to \cref{eq:betaThreshold}, is valid as long as $\beta \gg (\ell \omega)^{-1/6}$ (generally expected at large frequencies). 
Moreover, assuming for example that $\abs{\vb{a}''} = \abs{\vb{b}''}$ at the cusp, we obtain
\begin{equation}
    \dv{P}{\omega} \approx G \mu^2 \frac{1}{\ell^{1/3}\omega^{4/3}} \frac{1 + (\cos\beta)/2}{\sin\beta}.
\end{equation}
This result, although not valid for $\beta$ arbitrarily small, shows indeed an enhancement of the radiated power due to the larger beam width occurring when $\beta$ gets close to $0$ or $\pi$.

When $\beta \ll (\ell \omega)^{-1/6}$, the beam aperture saturates when $\varphi \to 0$ or $\pi$, as shown in \cref{eq:deltaThetaBetaZero} (and in more detail in Appendix~\ref{app:demo}).
Then the beam can approximated by 
\begin{equation}
    \label{eq:tired}\Delta\theta(\varphi) \approx
    \begin{cases}
        \dfrac{(\ell\omega)^{-1/3}}{\abs{\sin\varphi}},
        & (\ell\omega)^{-1/6} < \varphi < \pi-(\ell\omega)^{-1/6},\\[6pt]
        (\ell\omega)^{-1/6},
        & \text{otherwise}.
    \end{cases}
\end{equation}
This corresponds to an  elongated diamond approximated by two horizontal lines (first line of \cref{eq:tired}) and two portions of circle (second line, see \cref{eq:deltaThetaBetaZero}). 
Then the radiated power per unit frequency reads
\begin{equation} \label{eq:dpdwEnhanced}
    \dv{P}{\omega} \approx G \mu^2 \frac{1}{\ell^{1/6} \omega^{7/6}}
\end{equation}
which is indeed less suppressed than the usual formula at large frequencies.

To see if this effect might have an impact for e.g.\ on the SGWB, one must study the statistics of cusp properties, and in particular of the angle $\beta$. 
Before moving on to this discussion in \cref{s:stats}, let us close this section by considering some special cusps for which the above developments need to be adapted.

\begin{figure*}[t]
    \centering
    \subfloat{
        \includegraphics[width=.33\textwidth]{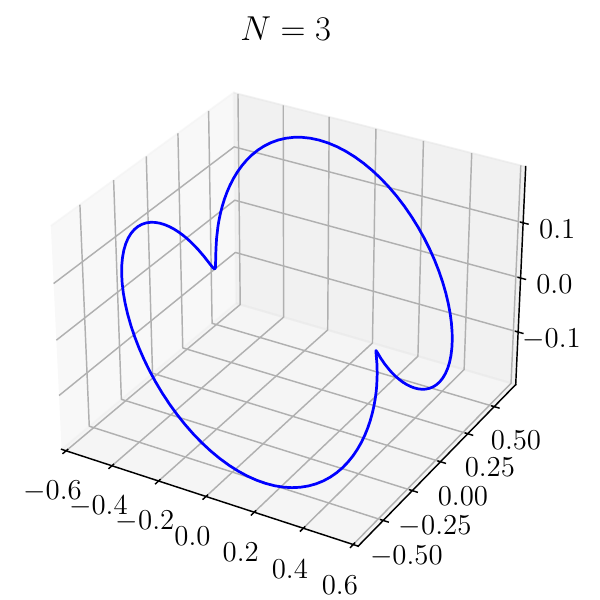}
    }
    \subfloat{
        \includegraphics[width=.33\textwidth]{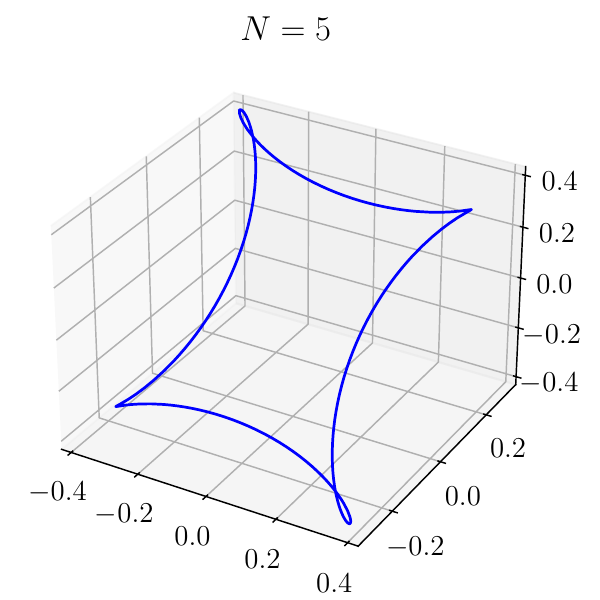}
    }
    \subfloat{
        \includegraphics[width=.33\textwidth]{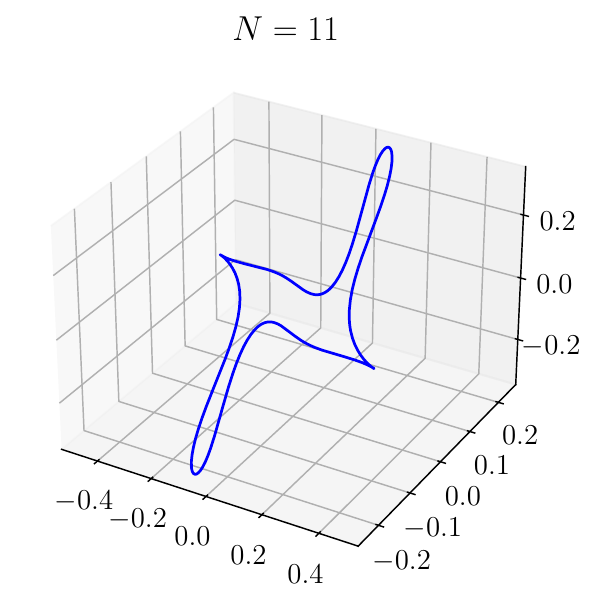}
    }
    \caption{Examples of odd harmonic loops for various values of the maximal harmonic order $N$. The snapshots are taken at a cusp time.}
    \label{fig:loopExamples}
\end{figure*}

\subsection{Flat cusps} \label{s:FlatCusps}

So far, we have assumed that $\vb{a}''$ and $\vb{b}''$, do not vanish at the cusp.
One might expect this to be justified, as their vanishing would appear to require some fine-tuning at the intersection of the two curves $\mathcal{C}_{\vb{a'}}$ and $\mathcal{C}_{\vb{b'}}$ precisely at the cusp. 
Recently, however, the amplitude of the GW radiated by cusps with vanishing derivatives has been studied in Ref.~\cite{Drew:2025txu}, highlighting some distinctive signatures. We outline these and extend the calculation of \cite{Drew:2025txu} to discuss the GW power emitted.

Consider `generalized cusps' described by 2 integers $(n,m)$ such that
\begin{equation}
    \forall k < n,  \,
    \abs{\vb{a}^{(k)}(u_c)} = 0 \text{ and }  \abs{\vb{a}^{(n)}(u_c)} \neq 0
\end{equation}
and 
\begin{equation}
    \forall k < m, \,
    \abs{\vb{b}^{(k)}(v_c)} = 0 \text{ and }  \abs{\vb{b}^{(m)}(v_c)} \neq 0
\end{equation}
The frequency dependence of the GW strain in the direction of the cusp is then given by~\cite{Drew:2025txu} 
\begin{equation}
\label{eq:drew-amplitude}
    h_{ij}^{\mathrm{TT}} \propto \frac{1}{\omega^{\beta_n + \beta_m}}, \quad
    \beta_n \equiv \frac{n}{2n-1} \, .
\end{equation}
As stated by the authors, \cref{eq:drew-amplitude} shows that cusps do not always radiate with a $\omega^{-4/3}$ spectrum,
\footnote{Standard cusps have $n = m = 2$ for which $\beta_n = \beta_m = 2/3$ so that the GW strain at the cusp of course scales as $\omega^{-4/3}$.}
but that this can flatten to $\omega^{-1}$ for large $n,m$. This leads to two questions.
First, how generic are these cusps? 
As suggested above these cases appear fine-tuned; this statement is made more quantitative in \cref{s:stats}.
Second, does the dependence of the radiated power per unit frequency, integrated over the whole beam, change for these special cusps? 
We now attempt to answer this second question. 

To get an order of magnitude estimate, we ignore the anisotropy of the beam-width discussed above, and simply approximate the beam as a cone (as in Ref.~\cite{Damour:2000wa})
whose opening angle we now calculate. 
The phase $p\phi(u)$ can be written as~\cite{Drew:2025txu}
\begin{equation}
\begin{split}
    p \, \phi(u) = &\frac{p}{2} \theta^2 u + \ldots \\&+\frac{p}{2} \frac{1}{(2n-1) [(n-1)!]^2} \abs{\vb{a}^{(n)} (u_c)}^2 u^{2n-1}
\end{split}
\end{equation}
where $\theta$ is the small angle between the velocity of the cusp and the direction of the observer.
Then the typical beam width can be estimated by equating the two terms above with $1$ \cite{Damour:2000wa}, yielding (here we assume $n=m$ for simplicity)
\begin{equation}
    \Delta \theta \propto \qty(\frac{1}{p})^{\frac{n-1}{2n-1}}.
\end{equation}
For example when $n=2$ (the standard case) $\Delta \theta \propto p^{-1/3}$ whereas for large $n$, $\Delta\theta \propto p^{-1/2}$. 
Hence, in contrast to the amplitude (which increases with increasing $n$), the beam aperture  narrows with increasing $n$, making these events even more difficult to observe.

Following the same procedure as in the preceding section, see \cref{eq:dpdwWithAngle}, the power per unit frequency can be estimated as
\begin{equation}
    \dv{P}{\omega} \propto \omega^2 h^\text{TT}_{ij}(\omega) h^\text{TT}_{ij}(\omega) \Delta \theta^2
\end{equation}
which yields an overall frequency dependence
\begin{equation} \label{eq:dpdwFlat}
    \dv{P}{\omega} \propto \omega^{-\frac{2n}{2n-1}} =
    \begin{cases}
        \omega^{-4/3} & \text{if } n=2 \\
        \omega^{-6/5} & \text{if } n=3 \\
        \omega^{-8/7} & \text{if } n=4 \\
    \end{cases}
\end{equation}
Hence, the relative growth of the amplitude wins over the narrowing of the aperture, so that the power per unit frequency becomes actually less suppressed at high frequencies. 
This is an important result, as the exponent in the dependence of the power per harmonic as a function of the harmonic number is a key parameter for predicting SGWB spectra~\cite{Blanco-Pillado:2017oxo,Ringeval:2017eww}. 
Hence, one should answer the question of the generality of such special cusps. To do so we now turn to a statistical study of cusp properties.

\section{Statistical study of cusp properties}\label{s:stats}

\subsection{Odd harmonic loops}

As discussed in detail in the previous sections, 
the properties of GWs radiated by a cusp are determined by its geometric features.
In particular, we found two unusual cases that can lead to an enhancement of the radiated gravitational power: either $\beta \to 0$ (in \cref{s:beamShape}) or flat cusps (in \cref{s:FlatCusps}).
In this section, building upon a method laid out in Ref.~\cite{Pazouli:2020qmj}, we present a statistical analysis of these key geometric parameters, in order to evaluate how generic these enhancement effects might be.

For this analysis, we rely on a large class of analytic loop solutions known as \emph{odd harmonic loops}~\cite{Siemens:1994ir}. 
For these, the $2\pi$-periodic $\vb{a}'$ and $\vb{b}'$ are written as finite Fourier sums of maximal harmonic order $N$, and the Fourier series only feature odd terms.
After removing redundant parameters, it can be shown~\cite{Siemens:1994ir} that $\vb{a}'$ is determined by $(N-1)/2$ polar angles $\theta_n^a \in [0, \pi]$ and $(N+1)/2$ azimuthal angles $\varphi_n^a \in [0, 2\pi]$, whereas $\vb{b}'$ is characterized by $(N-1)/2$ polar and azimuthal angles $\theta_n^b$ and $\varphi_n^b$. 
The Fourier coefficients for evaluating $\vb{a}'$ and $\vb{b}'$ are then obtained through the recurrence relations presented in Ref.~\cite{Brown1991}. 
Examples of odd harmonic loops with various values of the maximal harmonic order $N$ are displayed \cref{fig:loopExamples}.
Odd harmonic loops always have at least $2$ cusps as we show in Appendix~\ref{s:proofCuspsOddLoops}.

To study the geometric properties of cusps in the case of odd harmonic loops, we proceed as follows. 
First, we sample a particular loop with a uniform sampling on the sphere of the angles $(\theta_n^{a/b}, \varphi_n^{a/b})$ 
Then, we numerically identify cusp coordinates $(u_c, v_c)$ by looking for intersections of $u \mapsto \vb{a}'(u)$ and $v \mapsto \vb{b}'(v)$ on the Kibble-Turok sphere. 
Finally, we obtain the geometric properties of the cusp by evaluating the second derivatives $\vb{a}''(u_c)$ and $\vb{b}''(v_c)$. 
Details on the methods used for sampling and finding intersections, as well as comparisons of our results with the ones of Ref.~\cite{Pazouli:2020qmj} are found in Appendix~\ref{app:OddHarmonicLoops}.

A limitation of this approach is that we restrict ourselves to a subclass of all possible loops. 
Nonetheless, this subclass allows us to sample an arbitrary number of loops displaying a wide variety of cusp configurations.
In principle, as noted in Refs.~\cite{Allen:1990tv,Siemens:1994ir}, loops produced by self-intersections of larger loops, or infinite strings, tend to display high harmonic components (e.g.\ kinks). 
Nonetheless, as shown in Ref.~\cite{Siemens:1994ir}, high harmonic loops tend to self-intersect and fragment to lower harmonic loops and, as studied in Ref.~\cite{Pazouli:2020qmj} and Ref.~\cite{Pazouli:2021orp}, the GW predictions made by only considering low harmonics appear quite robust to the introduction of higher harmonic components. 
Hence, it is likely that physical loops could be well represented by low $N$ odd harmonic loops, with uniformly sampled angular parameters.

\begin{figure*}[t]
    \centering
    \subfloat{
        \includegraphics[width=.45\textwidth]{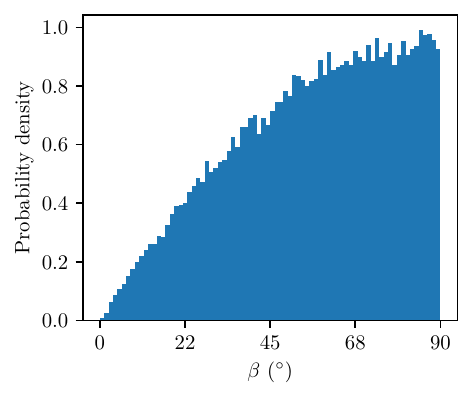}
    }
    \hfill
    \subfloat{
        \includegraphics[width=.45\textwidth]{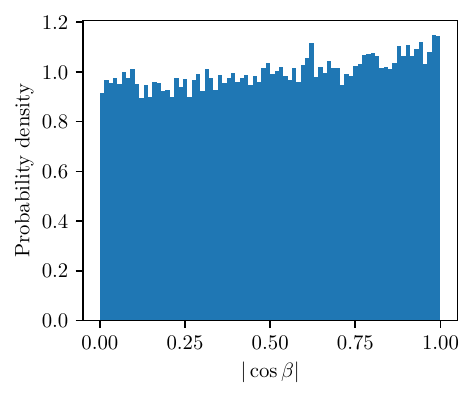}
    }
    \caption{Probability density of the angle of intersection $\beta$, obtained from $10~000$ realizations of odd harmonic loops of order $N=5$, with $10$ cusps per loop on average.
    }
    \label{fig:histo_beta}
\end{figure*}

\subsection{Results}

\subsubsection{Angle of intersection on the Kibble-Turok sphere}

\cref{fig:histo_beta} shows the distribution of the angle $\beta$ between 
$\vb{a}'$ and $\vb{b}'$ at the cusp.
The left panel shows that $\beta$ is more likely to be close to $90^\circ$. 
This was to be expected, as intersections of curves on the Kibble-Turok sphere tend to be more orthogonal than tangent.
The right panel shows that the distribution of $\abs{\cos\beta}$ can be approximated by a uniform distribution. 
Thus one can estimate easily the fraction of cusps that could present an enhancement due to sufficient alignment of $\vb{a}''$ and $\vb{b}''$. 
Following \cref{s:RadiatedPower}, the threshold $\beta_{\mathrm{th}}(\omega)$ is given in \cref{eq:betaThreshold} namely
\begin{equation}
    \beta_{\text{th}} \propto (\ell \omega)^{-1/6} \, .
\end{equation}
Due to the uniform distribution in $\cos \beta$, the fraction of cusps with enhanced beam width at $\omega$ can be estimated by
\begin{equation}
    \Pr(\beta \leq \beta_{\mathrm{th}}) = A (\ell \omega)^{-1/3} \, ,
\end{equation}
with $A$ a normalization constant.
This last result, combined with \cref{eq:dpdwusual,eq:dpdwEnhanced} enables us to estimate the typical radiated power per unit frequency as
\begin{multline}
    \dv{P}{\omega} \approx G\mu^2 \left(\frac{1}{\ell^{1/6} \omega^{7/6}}\times \frac{A}{\ell^{1/3}\omega^{1/3}}\right. \\ +\left. \frac{1}{\ell^{1/3}\omega^{4/3}} \times \qty[1-\frac{A}{\ell^{1/3}\omega^{1/3}}]\right) \, .
    \label{eq:flatrhombus}
\end{multline}
Overall, the radiated power of our population of loops is dominated by
\begin{equation}
    \dv{P}{\omega} \approx \frac{G \mu^2}{\ell^{1/3} \omega^{4/3}} + \order{\omega^{-3/2}} \, .
\end{equation}
Hence, even though aperture-enhanced cusps radiate more gravitational power, their relative scarcity suppresses their contribution and the traditional result of \cref{eq:dpdwusual} is robust under more precise considerations regarding the beam width.

\subsubsection{Vanishing of the second derivatives of the loop halves}

The magnitude of the loop halves second derivatives is usually characterized by two numbers 
\begin{equation}
    g_1 = \qty( \abs{\vb{a}''(u_c)} \abs{\vb{b}''(v_c)})^{-1/3}
\end{equation}
and 
\begin{equation}
    g_2 = \qty( \min \abs{\vb{a}''(u_c)},\abs{\vb{b}''(u_c)})^{-1} \, .
\end{equation}
The gravitational wave amplitude is proportional to $g_1$ (see \cref{eq:httx}) while the traditional beam aperture of \cref{eq:usualBeamWidth} can be shown to be proportional to $g_2^{-1/3}$~\cite{Damour:2000wa}. 

We are now interested in studying whether $g_{1,2}$ can display very large values, which would indicate that the flat cusps developed in Ref.~\cite{Drew:2023tob} and presented in \cref{s:FlatCusps} might be frequent events.

\begin{figure}[h]
    \centering
    \includegraphics{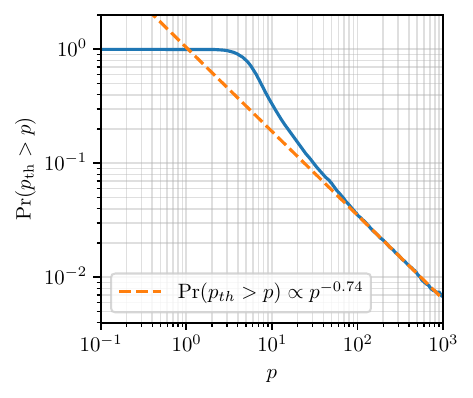}
    \caption{Complementary cumulative distribution function (blue) of $p_\text{th}$ obtained from $1~000$ odd harmonic loops of order $N=11$ with $22$ cusps per loop on average. 
    The large $p$ region is fit with a power law (orange).}
    \label{fig:histo_pth}
\end{figure}

The authors of Ref.~\cite{Pazouli:2020qmj} already presented histograms of $g_1$ and $g_2$ that do not seem to feature particularly large values of these parameters. 
Our findings, presented in \cref{app:OddHarmonicLoops}, corroborate this result. 
Hence, at least for odd harmonic loops, it seems very unlikely that an intersection of derivatives of the loop halves on the Kibble-Turok sphere concurs with a significant suppression of the loop halves second derivatives.
However, it is still possible that the second derivative at the cusp is small enough, compared to higher order terms, so that the usual saddle point approximation is not numerically accurate at some fixed high frequency. 
More precisely, considering a direction of the observer aligned with the velocity of the cusp, the phase $\phi(u)$ given in \cref{eq:phi_of_u} can be expanded near the cusp $u_c$ as 
\begin{multline}
    p \phi(u) = p\left\{\frac{1}{6}\qty[\vb{a}''(u_c)]^2 (u-u_c)^3\right. \\  + \left. \frac{1}{40} \qty[\vb{a}^{(3)}(u_c)]^2 (u-u_c)^5 + \mathcal{O}\qty[(u-u_c)^7]\right\}.
\end{multline}
The question we want to answer is whether, for a fixed frequency $p$, the integral is dominated by the term of order $(u-u_c)^3$ or $(u-u_c)^5$.

At arbitrarily large frequencies, the dominant behavior of the integral $\vb{I}_+$ is given by the first term above. 
At fixed cusp geometry, we can estimate the frequency above which the saddle point becomes accurate $p_{\mathrm{th}}$ by equating these terms to $1$:
\begin{equation}
    p \geq p_{\mathrm{th}} = \frac{\qty{\qty[\vb{a}^{(3)}(u_c)]^2}^{3/2}}{\qty{\qty[\vb{a}''(u_c)]^2}^{5/2}}.
\end{equation}
If the second derivative is too small, one has to go to larger harmonics for the saddle point estimation to be correct.

Therefore, we can estimate the proportion of loops that will feature flat cusps by evaluating the complementary cumulative distribution function $\Pr(p_\mathrm{th} > p)$ for odd harmonic loops. 
This function exactly represents the proportion of loops such that, at a fixed frequency $p$, the usual dominant order saddle point estimation cannot be trusted.

\cref{fig:histo_pth} shows this complementary cumulative distribution function of $p_\mathrm{th}$ obtained from the statistical analysis of $10~000$ realizations of odd harmonic loops of order $N=3$. 
A first observation is that $p_\mathrm{th}$ is typically of order unity, which is expected as it is the ratio of two dimensionless numbers. 
However, we also notice a fat tail that we estimate to scale as
\begin{equation} 
    \label{eq:proportionFlatCusps}
    \Pr(p_\mathrm{th} > p) = B \, p^{-0.74} \, ,
\end{equation}
with $B$ a normalization constant, for a population of loops with harmonic order $N=11$.

Similarly to the arguments around \cref{eq:flatrhombus}, we can estimate the typical radiated power per unit frequency of our population of loops with derivative order $n=3$, taking into account the occurrence of flat cusps and their radiated power \cref{eq:dpdwFlat}.
\begin{multline}
    \dv{P}{\omega} \approx G\mu^2 \left(\frac{1}{\ell^{1/6} \omega^{6/5}}\times \frac{B}{\ell^{0.74}\omega^{0.74}}\right. \\ +\left. \frac{1}{\ell^{1/3}\omega^{4/3}} \times \qty[1-\frac{B}{\ell^{0.74}\omega^{0.74}}]\right) \, .
\end{multline}
Overall, the radiated power is dominated by
\begin{equation}
    \dv{P}{\omega} \approx \frac{G \mu^2}{\ell^{1/3} \omega^{4/3}} + \order{\omega^{-1.94}} \, .
\end{equation}
The scarcity of flat cusps makes their contribution subdominant in the radiated power per unit frequency, even though they radiate more power.

\section{Conclusion}\label{s:conclusion}

We have revisited the seminal computation of the gravitational wave (GW) emission from a cosmic string cusp of Refs.~\cite{Damour:2000wa,Damour_2001}.
We identified two distinctive features that, to the best of our knowledge, were not discussed in the literature: at any given frequency, the GW emission of cusps is not concentrated in a cone but instead in a diamond; and the maximum of the emission is localized in two lumps away from the direction of the cusp.
To understand these results, we carried an in-depth analysis of the GW emission of a cusp, including terms that were neglected in Refs.~\cite{Damour:2000wa,Damour_2001}.

First, the diamond shape appears because two preferred directions, namely the second derivative of the loop halves $\vb{a}''$ and $\vb{b}''$, break the symmetry.
In particular, we showed that the beam aperture can be written as (see \cref{eq:beamWidth})
\begin{equation}
    \Delta \theta(\varphi) \approx \frac{2}{p^{1/3}} \qty[\frac{\abs{\sin\varphi}^3}{\abs{\vb{a}''(u_c)}} + \frac{\abs{\sin(\varphi-\beta)}^3}{\abs{\vb{b}''(v_c)}}]^{-1/3} .
\end{equation}
Naturally, one may ask what happens when the two directions become degenerate, i.e.\ when $\varphi=\beta=0$. 
The emission then becomes enlarged in one direction (see \cref{eq:deltaThetaBetaZero}), a rare event that may nonetheless be of importance for future burst searches.
We showed that the overall effect on total radiated power is substantially enhanced for these configurations.

Second, the direction of maximum GW emission at any given frequency is misaligned with the cusp velocity because of interferences in the transverse and traceless projection.
The authors in Refs.~\cite{Damour:2000wa,Damour_2001} had shown that the first order contribution of the $\vb{I}_\pm$ is longitudinal and is removed by the TT gauge, hence not physical.
This is no longer true when the emission of the GW is not the direction of the cusp velocity and the first order contribution of $\vb{I}_\pm$ leaks in the source term of the GW.
We have calculated this effect analytically for a simple loop and shown that the emission becomes maximal along a preferred axis.
This effect is only of order unity, which may explain why it remained unnoticed by numerical works using logarithmic scales.
Note also that, for the same reason, the GW is not linearly polarized away from the direction of the cusp velocity.
In frequency space, this enhancement could manifest itself as an order unity bump in the waveform, at the frequency at which the direction of the maximum and the observer are aligned.
Perhaps, a more intuitive interpretation for these two phenomena is that the two ``legs'' of the cusp break the circular symmetry around the cusp. 
These legs are also moving at relativistic velocities and contribute to the overall emission of GWs but in slightly misaligned directions.

Finally, we extended our study of the beam aperture to the flat cusps introduced in Ref.~\cite{Drew:2025txu}, which have vanishing higher-order derivatives.

To evaluate the overall effect of our findings in a generic population of cosmic string loops, we performed a statistical study in a controlled subpopulation: the odd-harmonic loops.
Overall, we find that the configurations having an enhanced emission of GW are too rare events to have an effect on the average radiated power by the loop population. 

In conclusion, we do not expect a particular impact on averaged quantities such as the cosmic string loop population\footnote{Standard models of cosmic string loop population include a single parameter $\Gamma$ for the averaged power emitted in GWs.} and the SGWB. 
However, the occurrence of these new configurations follows a ``fat'' power-law distribution.
Therefore, they might be relevant for any observable that depends on rare events, such as direct burst searches in ground-based detectors and LISA.

We stress that our analysis is supported by a statistical analysis of odd-harmonic loops which may not represent accurately physical loops, and we would need a similar analysis on a more physical sample of loops to have a definitive conclusion.
Additionally, it remains to be seen how this analysis translates to the GW emission by kinks and the quantitative impact on GW searches.

\begin{acknowledgments}
    We thank Christophe Ringeval and Tanmay Vachaspati for comments in the early stages of this manuscript, and Amelia Drew for useful discussions. We also acknowledge the very fruitful environment and discussions at the workshop ``Universe Unravelled: Cosmic strings at 50''  in Cambridge, september 2026. DAS is grateful to CERN for hospitality during the final stages of this paper.
\end{acknowledgments}

\section*{AI disclosure}
No generative AI was used during the preparation of this manuscript. 
The text, codes and figures were made with love by humans for humans.
In these trying times, please cherish genuine human-made art and engage in passionate conversations with your fellow human colleagues.

\appendix

\section{Saddle point approximation} \label{s:saddlePoint}

We explain how the saddle point approximation used for establishing \cref{eq:decayI} is carried out. 
The aim is to estimate 
\begin{equation}
    \vb{I}_+(p, \hat{n}) = \frac{\ell}{2\pi} \int_{-\pi}^\pi \vb{a}'(u) e^{i p \phi(u)} \dd{u}
\end{equation}
in the large frequency $p\gg1$ regime and with
\begin{equation}
    \phi'(u) = 1-\vb{a}'(u) \cdot \hat{n}
\end{equation}
when there is a complex root given, to first order in $\delta_a$, by 
\begin{equation}
    u^\star = u_0 + i \frac{\delta_a}{\abs{\vb{a}''(u_0)}}.
\end{equation}

Let us write $\epsilon = \Im(u^\star)$ and deform the integration contour $[-\pi, \pi]$ into 
\begin{equation}
    -\pi \rightarrow -\pi  +i\epsilon \rightarrow \pi+i\epsilon \rightarrow \pi.
\end{equation}
In the center of mass frame, $u \mapsto \vb{a}(u)$ is $2\pi$-periodic. Since $p$ is integer, the integrals on the vertical parts of the contour cancel out. Therefore, 
\begin{equation}
    \vb{I}(p, \hat{n}) = \frac{\ell}{2\pi} \int_{-\pi+i\epsilon}^{\pi + i\epsilon} \vb{a}'(u) e^{ip \phi(u)} \dd{u} \, .
\end{equation}
Now, this integral can be evaluated using the fact that $u^\star$ is a saddle point. 
For that, though, we need to ensure that the second derivative of $\phi$ has the right sign. Using \cref{eq:derivativesAtu0} and \cref{eq:derivativesAtu0Bis} we find
\begin{equation}
    \Im[\phi''(u^\star)] = \vb{a}''^2(u_0) \epsilon > 0,
\end{equation}
provided we considered the complex root having a positive imaginary part. Hence, the saddle-point approximation yields
\begin{multline}
    \vb{I}_+(p, \hat{n}) \underset{p\rightarrow\infty}{\sim} \frac{\ell}{2\pi} e^{-p\Im[\phi(u^\star)]} \\ \int_{-\infty}^{+\infty} \vb{g}(u) e^{-\frac{p}{2} \Im[\phi''(u^\star)](u-u^\star)^2} \dd{u}
\end{multline}
where $\vb{g}(u)$ is a bounded function that incorporates the contributions from the prefactor $\vb{a}'(u)$ and from the complex exponential $\exp{ip\Re[\phi(u)]}$. 

As a consequence, the first term $\exp{-p\Im[\phi(u^\star)]}$ determines the exponential decay of $\vb{I}_+$ in terms of $p$.

\section{Beam aperture in the general case}\label{app:demo}

Recall that the beam aperture is given by the distances $\delta_a$ and $\delta_b$ between the observer's direction $\hat{n}$ and the two curves $\mathcal{C}_{\vb{a}'}$ and $\mathcal{C}_{\vb{a}'}$.
Indeed, \cref{eq:awful} reads
\begin{equation}
    \abs{\bar{h}_{ij}} \propto \exp\left\{-\frac{1}{2} \left[
    \qty(\frac{\delta_a}{\Delta_a(u_1, p)})^3 + \qty(\frac{\delta_b}{\Delta_b(v_1, p)})^3
    \right] \right\} \, .
\end{equation}
Begin with the part coming from $a$. The first step is to evaluate $\delta_a$ as a function of $\theta$ and $\varphi$. Near the cusp, the curve $\mathcal{C}_{\vb{a}'}$ is described by
\begin{equation}
    \vb{a}'(u) = \vb{a}'(u_c) + (u-u_c) \vb{a}''(u_c) + \frac{1}{2}(u-u_c)^2 \vb{a}^{(3)}(u_c) \, .
\end{equation}
We consider the plane $\Pi$ ($\hat{i}$, $\hat{j}$) tangent to the Kibble-Turok sphere as in \cref{eq:basis}. 
Projecting the above equation onto $\hat{i}$ and $\hat{j}$ yields the following parametric equation for this curve in the plane $\Pi$:
\begin{equation}
    \left\{ \begin{array}{ccc} 
        x &=& (u-u_c) \vb{a}''(u_c), \\
        y &=& \frac{1}{2} (\vb{a}^{(3)}(u_c) \cdot \hat{j}) (u-u_c)^2 \, .
    \end{array}  \right.
\end{equation}
Hence, the curve $\mathcal{C}_{\vb{a}'}$ is a parabola described by
\begin{equation}
    y(x) = q_a x^2 \quad \text{with} \quad q_a = \frac{\vb{a}^{(3)}(u_c) \cdot \hat{j}}{2 \abs{\vb{a}''(u_c)}^2}.
\end{equation}
In the plane $\Pi$, the direction to the observer (its projection to the plane) is given by 
\begin{equation}
    \hat{n} = \theta \cos \varphi~\hat{i} + \theta \sin\varphi~\hat{j}.
\end{equation}
Hence, the (squared) distance between $\hat{n}$ and a point parametrized by $u$ (or $x(u)$) on the curve $\mathcal{C}_{\vb{a}'}$ is 
\begin{equation}
    d_a(x)^2 = (x - \theta \cos\varphi)^2 + \qty[q_a x^2 - \theta \sin \varphi]^2.
\end{equation}
The coordinate $x_a$ corresponding to the minimal distance is given by solving 
\begin{equation}
    \dv{d_a^2}{x}~(x_a) = 0.
\end{equation}
This is a polynomial of degree $3$, but we can find an expansion for $x_a$ using the smallness of $\theta$,
\begin{equation}
    x_a = \theta \cos \varphi + 2 q_a \cos \varphi \sin \varphi \theta^2 + \mathcal{O}(\theta^3) \, .
\end{equation}
From this, the distance between $\hat{n}$ and $\mathcal{C}_{\vb{a}'}$ is given by
\begin{multline}
    \delta_a^2 \equiv d_a(x_a)^2 = \theta^2 \sin^2 \varphi + (-2 q_a \theta^3 \cos^2 \varphi \sin \varphi) \\
    + \theta^4  q_a^2 \cos^4 \varphi + (-\theta^4 c^2 \cos^2 \varphi \sin^2 \varphi) + \order{\theta^5} \, .
\end{multline}
There are two main regimes depending upon $\varphi$.
If $\sin\varphi$ is not too small, then the first term of the expansion dominates.
On the contrary, if $\varphi$ is too close to $0$ or $\pi$ then $\sin \phi \to 0$, and all the terms vanish except the third. 

Overall, a good approximation of $\delta_a$ that captures correctly both regimes is given by
\begin{equation}
    \delta_a^2 \approx \theta^2 \sin^2 \varphi  
    + q_a^2 \theta^4 \cos^4 \varphi \, ,
\end{equation}
which corresponds to \cref{eq:delta_a_full} in the main text.

Then, we need to consider $\Delta_a$, which we can immediately expand in $\theta \ll 1$ 
\begin{equation}
    \qty(\frac{1}{\Delta_a(u_1, p)})^3 = \frac{p}{\abs{\vb{a}''(u_c)} + \mathcal{O}(\theta)} = \frac{p [1 + \mathcal{O}(\theta)]}{\abs{\vb{a}''(u_c)}} \, .
\end{equation}
Therefore, the contribution to the decay from $\vb{I}_+$ is
\begin{multline}
    \exp{\frac{-1}{2} \frac{\delta_a^3}{\Delta_a(u_1, p)^3}} = \exp 
    \left\{\frac{-p}{2 \abs{\vb{a}''(u_c)}} \left[\theta^2 \sin^2 \varphi \right.\right.\\
    \left. \left. + \# + \theta^4 q_a^2 \cos^4 \varphi + \order{\theta^5}\right]^{3/2}  \right\}
\end{multline}
where the $\#$ denotes a term that is subdominant (in $\theta$) when $\sin\varphi \neq 0$ and that vanishes at $\sin\varphi=0$. 
Thus there are two regimes yielding either $\Delta \theta \propto p^{-1/3}$ or $\propto p^{-1/6}$. 

On combining with a similar term for $\delta_b$, the beam decay is fixed by setting the term in curly brackets to $1$ (to obtain the characteristic decay width), namely 
\begin{multline}
    p\left(\frac{\qty[\theta^2 \sin^2 \varphi + \theta^4 q_a^2 \cos^4 \varphi]^{3/2}}{2 \abs{\vb{a}''(u_c)}} \right. \\ + \left.\frac{\qty[\theta^2 \sin^2 (\varphi-\beta) + \theta^4 q_b^2 \cos^4 (\varphi-\beta)]^{3/2}}{2 \abs{\vb{b}''(v_c)}} \right)  = 1 \, .
    \label{eq:beam-all}
\end{multline}
All these contributions add positively, therefore the behavior of the beam aperture is determined by the dominant of these four terms.
When these four terms contribute equally in \cref{eq:beam-all}, one finds in the limit $\varphi, \beta \ll 1$
\begin{equation}
    \begin{cases}
        q_a^2 \theta^2 = \varphi^2 \\
        q_b^2 \theta^2 = (\varphi - \beta)^2 \\
        2 p \varphi^6 = q_a^3 \abs{\vb{a}''(u_c)} \\
        2 p (\varphi - \beta)^6 = q_b^3 \abs{\vb{b}''(v_c)} \, .
    \end{cases}
\end{equation}
Therefore, the threshold value between $\Delta \theta \propto p^{-1/3}$ and $p^{-1/6}$ is
\begin{equation}
    \beta_\mathrm{th} \propto p^{-1/6} \, , 
\end{equation}
for values of $\varphi \ll p^{-1/6}$ as  reported in \cref{eq:betaThreshold}

\section{Deviation from the calculation by Damour and Vilenkin}
\label{app:deviation}
Here we explain the difference between our method and the one used in Ref.~\cite{Damour_2001} for estimating the beam aperture.

Using our notation, the calculation of Damour and Vilenkin can be explained as follows. Considering an angle $\theta$ between the cusp (located at $u = 0$), the phase $\phi(u)$ can be expanded near the cusp as 
\begin{equation}
    \phi(u) = \frac{\theta^2}{2}u + \frac{\theta \cos\alpha}{2} \abs{\vb{a}''(0)} u^2 + \frac{1}{6} \abs{\vb{a}''(0)}^2u^3 
\end{equation}
where $\alpha$ is the angle between $\hat{n}_c - \hat{n}$ and $\vb{a}''(0)$. Here, one recovers the fact there is no real saddle point, as the discriminant of the derivative of the above phase is negative. To estimate the exponential decay, Damour and Vilenkin equate the first and the last term of the above equation to $1/p$, to find the typical angle $\theta$ that separates the domination of the last term (which is the result as $\theta = 0$), and that of the first term (which accounts for the exponential suppression in $\theta$ at large $\theta$). 

Neglecting the term in $\alpha$ is precisely what gives an isotropic result. 
In Ref.~\cite{Damour_2001}, the authors discard the term in $\alpha$ which amounts to taking $\cos \alpha = 0$.
In this special case, the roots of $\phi'(u)$ are trivial and $\theta = \delta$.
In the present article, we discuss precisely this dependence on $\alpha$ and some of the arguments of Ref.~\cite{Damour_2001} cannot be directly applied. 
In particular, $\delta \neq \theta$ as can be seen in \cref{fig:delta_a_b} and as we explicit in \cref{s:beamShape}.

\section{Numerical methods for studying odd harmonic loop cusps} \label{app:OddHarmonicLoops}

\subsection{Product representation}

We decompose the two loop halves in Fourier series and keep only a finite number of harmonics
\begin{equation}
    \vb{a}'(u) = \vb{Z} + \sum_{n=1}^N  \qty(\vb{A}_n \cos n u + \vb{B}_n \sin n u)
\end{equation}
and
\begin{equation}
    \vb{b}'(v) = -\vb{Z} + \sum_{n=1}^N  \qty(\mathcal{C}_n \cos n u + \vb{D}_n \sin n u),
\end{equation}
where the constraint on the $0$-th order terms stems from the spatial periodicity of the loop. 

The constraint on the norm $\norm{\vb{a}'} = \norm{\vb{b}'} = 1$ restricts the allowed values of the Fourier coefficients. 
A simple way to describe the most general solutions for $\vb{a}'$ and $\vb{b}'$ at a given order $N$ was presented in Ref.~\cite{Brown1991} and is known as the \emph{product representation}. 
The idea is to satisfy the norm constraint by defining $\vb{a}'$ and $\vb{b}'$ as resulting from compositions of $N$ rotations of a unit vector. 
More precisely, if $u \mapsto \vb{f}(u)$ is a function expressed as a Fourier sum of order $N$, such that $\vb{f}(u)^2=1$, then there exists $N+1$ pairs of angles $(\theta_n \in [0, \pi], \varphi_n \in [0, 2\pi])$ such that
\begin{equation} \label{eq:productRepresentation}
    \vb{f}(u) = \rho_{N+1} R_z(u) \rho_N R_z(u) \ldots \rho_3 R_z(u) \rho_2 R_z(u) \rho_1 \vu{z}
\end{equation} 
where
\begin{equation}
    \rho_n = \rho(\theta_n, \varphi_n) = R_z(-\theta_n) R_x(\varphi_n) R_z(\theta_n),
\end{equation}
$\vu{z}$ is the unit vector at the north pole of the unit sphere and $R_x$ and $R_z$ are rotation matrices around the $x$ and $z$ axis respectively.

\cref{eq:productRepresentation} describes all possible vectors.
First, observe that there are $N$ rotations involving the variable $u$, so at most a term in $(\cos u)^N$ and hence at most the harmonic of order $N$. Second, the condition on the norm forces $4N+1$ constraints on the $6N+3$ Fourier coefficients, which matches the $2(N+1)$ angles from this last formulation. 
A rigorous derivation is given in Ref.~\cite{Brown1991}.

In the center of mass frame
\begin{equation}
    \int_0^\ell \dd{\sigma} \pdv{\vb{X}}{t} = 0 \, ,
\end{equation}
and the functions $\vb{a}$ and $\vb{b}$  also become $2\pi$-periodic, meaning that $\vb{Z} = \vb{0}$. 
However, if one wishes to use the product representation, it is not clear how to choose the parameters $(\theta_n , \varphi_n)$ in order to satisfy this constraint~\cite{Siemens:1994ir,Pazouli:2021orp}. 
A solution is to restrict attention to a smaller class of solutions containing only odd harmonics (meaning $N$ is also odd), which enables one to automatically satisfy the center-of-mass constraint. 
Then, after removing redundant angles, the loop halves can be parametrized as~\cite{Siemens:1994ir}
\begin{multline}
    \vb{a}'_N(u) = R_x(\tilde{\varphi}^a_{N+2}) R_z(2u) \rho(\tilde{\theta}_N^a, \tilde{\varphi^a_{N}}) \\ \ldots R_z(2u) \rho(\tilde{\theta}_3^a, \tilde{\varphi}_3^a) R_z(u) \vu{x}
\end{multline}
and 
\begin{multline}
    \vb{b}'_N(v) = R_z(2v) \rho(\tilde{\theta}_N^b, \tilde{\varphi}^b_{N}) \\ \ldots R_z(2v) \rho(\tilde{\theta}_3^b, \tilde{\varphi}_3^{b}) R_z(v) \vu{x}
\end{multline}
where the product is only taken on odd indices $n$ and where the matching with the previous angles $(\theta_n, \varphi_n)$ is given by
\begin{equation}
    (\theta_n, \varphi_n) = \left\{ \begin{array}{ccc }
        (-\pi/2 , \pi/2) & \text{ if } & n=1, \\
        (0, 0) & \text{ if } & n \geq 3 \text{ is odd}, \\
        (\tilde{\theta}_{n+1}, \tilde{\varphi}_{n+1}) & \text{ if } & n \text{ is even}.
    \end{array} \right.
\end{equation}
We can again check by counting the cosines of $u$ and $v$ that the harmonic order is indeed $N$. Moreover, since $\vb{a}_N'(u+\pi) = - \vb{a}_N'(u)$ and $\vb{b}_N'(v+\pi) = - \vb{b}_N'(v)$, these expressions do indeed represent Fourier sums with only odd harmonics. It should be noted that $\varphi_1 = -\pi/2$ is also possible for having a vanishing average.

These expressions allow us to sample a large class of cosmic string loops. However, in order to achieve a computationally tractable sampling at large harmonic order $N$, it is preferable to work with Fourier sums. Fortunately, recurrence relations were provided in Ref.~\cite{Brown1991} to compute the Fourier coefficients from the angles.

\subsection{Recurrence relations on Fourier coefficients}

Let us consider a function $\mathcal{C}_N$ written as
\begin{equation} \label{eq:productRepresentationStandardForm}
    \mathcal{C}_N(u) = R_z(u) \rho_N R_z(u) \ldots \rho_3 R_z(u) \rho_2 R_z(u) \rho_1 \vu{z}
\end{equation} 
where
\begin{equation}
    \rho_n = \rho(\theta_n, \varphi_n) = R_z(-\theta_n) R_x(\varphi_n) R_z(\theta_n)
\end{equation}
$\vu{z}$ is the unit vector at the north pole of the unit sphere and $R_x$ and $R_z$ are rotation matrices.

Importantly, note that there is a slight difference between the first term in the right-hand side of \cref{eq:productRepresentation} and of \cref{eq:productRepresentationStandardForm}. 
This latter formulation is referred to as the ``standard form'' in Ref.~\cite{Brown1991}. 
In the end, the Fourier coefficients obtained with the method described below should be multiplied by $\rho_{N+1}$.

The goal is to find the Fourier coefficients $\mathcal{C}_n^N \in \mathbb{C}^3$ for some value of $N \geq 1$ and for all $n \in \{-N, \dots, N\}$ such that 
\begin{equation}
    \mathcal{C}_N(u) = \sum_{n=-N}^N \mathcal{C}_n^N e^{i n u}.
\end{equation}

The recurrence relations of Ref.~\cite{Brown1991} can be written as 
\begin{equation}
    \mathcal{C}_n^N = \vb{P} \mathcal{C}_{n+1}^{N-1} + \vb{Q} \mathcal{C}_n^{N-1} + \overline{\vb{P}} \mathcal{C}_{n-1}^{N-1}
\end{equation}
with the matrices
\begin{equation}
        \vb{P} = \begin{pmatrix}
        f_N & g_N & h_N \\
        if_N & ig_N & ih_N \\
        0 & 0 & 0
    \end{pmatrix}
\end{equation}
and
\begin{equation}
     \vb{Q} = \begin{pmatrix}
        0 & 0 & 0 \\
        0 & 0 & 0 \\
        \sin\varphi_N\sin\theta_N & \sin\varphi_N\cos\theta_N & \cos\varphi_N
    \end{pmatrix},
\end{equation}
where 
\begin{align}
    f_N &= \frac{1}{2} \qty[\cos^2\qty(\frac{\varphi_N}{2}) + \sin^2 \qty(\frac{\varphi_N}{2}) e^{i2\theta_N}], \\
    g_N&= \frac{-i}{2} \qty[\cos^2\qty(\frac{\varphi_N}{2}) - \sin^2 \qty(\frac{\varphi_N}{2}) e^{i2\theta_N}], \\
    h_N &= \frac{i}{2} \sin\varphi_N e^{i\theta_N},
\end{align}
and with the boundary conditions
\begin{equation}
    \mathcal{C}_0^0 = \begin{pmatrix}
        0 \\
        0 \\
        1
    \end{pmatrix} \quad \text{and} \quad \mathcal{C}_{N+1}^N = \mathcal{C}_{-(N+1)}^N = \begin{pmatrix}
        0 \\
        0 \\
        0
    \end{pmatrix}.
\end{equation}

Note that these recurrence relations highlight why it is difficult to find the set of angles $(\theta_n, \varphi_n)$ to go to the center of mass frame where $\mathcal{C}_0^N = \vb{0}$, as all of these angles are part of $\mathcal{C}_0^N$.

\subsection{Existence of cusps on odd harmonic loops} \label{s:proofCuspsOddLoops}

We show that odd harmonic loops always present cusps.
Consider two $2\pi$-periodic continuous functions $\vb{f}(u)$ and $\vb{g}(v)$ taking their values on the unit sphere $\mathcal{S}^2$, such that 
\begin{equation} \label{eq:symfg}
    \vb{f}(u + \pi) = -\vb{f}(u) \quad \text{and} \quad \vb{g}(v + \pi) = -\vb{g}(v).
\end{equation}
Let us also assume that $\vb{f}$ and $\vb{g}$ can be expanded in finite Fourier sums for simplicity.
Let us also define the parity operator
\begin{equation}
    \mathcal{P}: \vb{r} \mapsto -\vb{r}
\end{equation}
and let $\mathcal{C}_f$ and $\mathcal{C}_g$ respectively denote the closed curves by $\vb{f}$ and $\vb{g}$ on the unit sphere. 

By the property of \cref{eq:symfg}, $\mathcal{C}_f$ and $\mathcal{C}_g$ are invariant under $\mathcal{P}$. 
As $\vb{f}$ is a finite Fourier sum, its curve $\mathcal{C}_f$ defines a partition of $\mathcal{S}^2 \backslash \mathcal{C}_f$ in $M \geq 2$ disjoint domains $\qty(F_i)_{1 \leq i \leq M}$, because there can only be a finite number of self-intersections. For example, if $\mathcal{C}_f$ is a great circle, then $M=2$ and the domains are two hemispheres.

Importantly, as $\mathcal{C}_f$ is invariant under $\mathcal{P}$ and because of the continuity of $\vb{f}$, for each domain $F_i$ there exists an antipodal domain $F_j$ with $j \neq i$ such that $\mathcal{P}\qty(F_i) = F_j$.

Let $v \in [0, 2\pi]$. If $\vb{g}(v) \in \mathcal{C}_f$ then there exists an intersection between $\vb{f}$ and $\vb{g}$. If not, then there exists $i \in \{1, \dots, M\}$ such that $\vb{g}(v) \in F_i$. Hence $\mathcal{P}\qty[\vb{g}(v)] \in \mathcal{P}(F_i) = F_j$ for some $j \in \{1, \dots, M\}$ with $j \neq i$. But  $\mathcal{P}\qty[\vb{g}(v)] = -\vb{g}(v) = \vb{g}(v+\pi) \in \mathcal{C}_g$. Hence $\vb{g}(v)$ and $\vb{g}(v + \pi)$ belong to two distinct domains. A continuous curve that links these two points, such as $\mathcal{C}_g$, must therefore intersect $\mathcal{C}_f$.

\subsection{Finding cusps} \label{s:findingCusps}

In this context, cusps are found by solving
\begin{equation}
    \vb{a}'(u) = \vb{b}'(v).
\end{equation}
for $u$ and $v$. In the following, we focus on loop halves having the same maximal harmonic order $N$.

At large harmonic order $N$, there can be numerous solutions. 
The first step toward identifying all cusps is therefore to provide suitable initial guesses for a root-finding algorithm. 
To this end, we rely on a sampling of the loop halves on $N_u$ and $N_v$ points. 
Intuitively, this sampling must become finer at higher harmonic order $N$, since progressively smaller-scale features appear. 
More precisely, let $\vb{f} = \left[\vb{a}' \text{ or } \vb{b}'\right]$, and define
\begin{equation}
    L = \int_0^{2\pi} \dd{u} \abs{ \vb{f}'(u) }
\end{equation}
the length of the curve traced by the loop-half derivative on the unit sphere.
Requiring that the arc-length spacing $\delta L$ between two consecutive samples satisfies
\begin{equation}
    \frac{\delta L}{L} < \epsilon,
\end{equation}
with $\epsilon$ a tolerance parameter taken to be $10^{-2}$ in the code, and using the Cauchy–Schwarz inequality to bound the length by
\begin{equation}
L < 2 \pi N,
\end{equation}
it follows that the number of samples $N_u$ and $N_v$ must satisfy
\begin{equation}
    N_u,~N_v > \frac{\sqrt{N}}{\epsilon}.
\end{equation}
Once this sampling has been performed, cusp candidates $(u, v)$ satisfying $\abs{\vb{a}'(u) + \vb{b}'(v)} < 0.1$ are identified using an optimized nearest-neighbor search based on a \verb|KDTree|~\cite{KDTree}.
Cusp candidates are then used as initial guesses for a root-finding algorithm solving $\abs{\vb{a}'(u) + \vb{b}'(v)} = 0$.

The final step is to remove duplicates. 
This requires checking distances both in $(u, v)$ space since different candidates may converge to the same $(u, v)$ values and in the three-dimensional geometric space, as possible sub-periodicities of $2\pi/k$ with $k>1$ can lead to different $(u, v)$ representations of a single cusp.
 It is useful to note that, since 
\begin{equation} \label{eq:symOddHarmonic}
    \vb{a}'(u+\pi) = -\vb{a}'(u) \quad \text{and} \quad \vb{b}'(v+\pi) = -\vb{b}'(v)
\end{equation}
for odd harmonic loops, the number of cusps is always even.

We attempted to reproduce the results from Ref.~\cite{Pazouli:2020qmj} concerning the mean number of cusps $N_c$ per period as a function of the harmonic order $N$, as well as the distribution of $N_c$ for different values of $N$. 
We identified two discrepancies between our results and theirs.

First, in constructing histograms for the distribution, the authors of Ref.~\cite{Pazouli:2020qmj} uniformly sample the angles $\tilde{\theta}_n$ and $\tilde{\varphi}_n$ in $[0, \pi]$ and $[0, 2\pi]$, respectively. 
In contrast, we choose to sample the pairs $(\theta_n, \varphi_n)$ uniformly on the sphere. 
This has the advantage of reducing the weight of values of $\theta_n$ close to $0$, which correspond to nearly tangent loop-half derivatives for which intersections are more difficult to detect numerically. 
This uniform spherical sampling therefore reduces numerical artifacts.

Nevertheless, when using the sampling method of Ref.~\cite{Pazouli:2020qmj}, we are able to reproduce their results, except for the case $N = 19$. 
In this case, we suspect that their result may be affected by under sampling. 
The histogram indeed appears consistent with aliasing effects, which we can reproduce by decreasing the values of $N_u$ and $N_v$. 
Moreover, as already mentioned, the number of cusps must always be even.

\bibliographystyle{apsrev4-1}
\bibliography{biblio}

\end{document}